\documentclass[a4paper,11pt]{article}

\usepackage{jinstpub}
\usepackage{graphicx}
\usepackage{xcolor}
\usepackage{caption}
\usepackage{subcaption}
\usepackage{multirow}
\usepackage[numbers,sort&compress]{natbib}
\usepackage{verbatim}
\usepackage{textgreek}

\title{Check on the features of potted 20-inch PMTs with 1F3 electronics prototype at Pan-Asia}

\author[a,b]{Caimei Liu,}
\author[a,b]{Min Li,}
\author[a,b,1]{Zhimin Wang,\note{Corresponding author.}}
\author[a]{Jun Hu,}
\author[f]{Nikolay Anfimov,}
\author[a]{Lei Fan,}
\author[g]{Alberto Garfagnini,}
\author[h]{Guanghua Gong,}
\author[a]{Shaojing Hou,}
\author[g]{Beatrice Jelmini,}
\author[a]{Xiaolu Ji,}
\author[a]{Xiaoshan Jiang,}
\author[f]{Denis Korablev,}
\author[c]{Tobias Lachenmaier,}
\author[a]{Si Ma,}
\author[a]{Xiaoyan Ma,}
\author[a]{Zhe Ning,}
\author[f]{Alexander G. Olshevskiy,}
\author[a,b]{Zhaoyuan Peng,}
\author[a]{Zhonghua Qin,}
\author[g]{Andrea Serafini,}
\author[g]{Katharina von Sturm,}
\author[c]{Tobias Sterr,}
\author[a]{Yunhua Sun,}
\author[c]{Alexander Felix Tietzsch,}
\author[e]{Jun Wang,}
\author[e]{Wei Wang,}
\author[a,b]{Yangfu Wang,}
\author[a]{Kaile Wen,}
\author[d]{Bjoern Soenke Wonsak,}
\author[a]{Wan Xie,}
\author[a]{Meihang Xu,}
\author[a,b]{Xiongbo Yan,}
\author[i]{Yifan Yang,}
\author[e]{Rong Zhao,}
\author[a,b]{Tong Zhou,}
\author[a]{Kejun Zhu}

\affiliation[a]{Institute of High Energy Physics, Beijing 100049, China}
\affiliation[b]{University of Chinese Academy of Sciences, Beijing 100049, China}
\affiliation[c]{Eberhard Karls Universität Tübingen, Physikalisches Institut, Tübingen, Germany}
\affiliation[d]{Institute of Experimental Physics, University of Hamburg, Hamburg, Germany}
\affiliation[e]{Sun Yat-Sen University, Guangzhou, China}
\affiliation[f]{Joint Institute for Nuclear Research, Dubna, Russia}
\affiliation[g]{Dipartimento di Fisica e Astronomia dell'Universita' di Padova and INFN Sezione di Padova, Padova, Italy}
\affiliation[h]{Tsinghua University, Beijing, China}
\affiliation[i]{Université Libre de Bruxelles, Brussels, Belgium}
\emailAdd{wangzhm@ihep.ac.cn}

\abstract{
The Jiangmen underground neutrino observatory (JUNO) is a neutrino project with a 20-kton liquid scintillator detector located at 700-m underground. The large 20-inch PMTs are one of the crucial components of the JUNO experiment aiming to precision neutrino measurements with better than 3\% energy resolution at 1\,MeV. The excellent energy resolution and a large fiducial volume provide many exciting opportunities for addressing important topics in neutrino and astro-particle physics. With the container \#D at JUNO Pan-Asia PMT testing and potting station, the features of waterproof potted 20-inch PMTs were measured with JUNO 1F3 electronics prototype in waveform and charge, which are valuable for better understanding on the performance of the waterproof potted PMTs and the JUNO 1F3 electronics. In this paper, basic features of JUNO 1F3 electronics prototype run at Pan-Asia will be introduced, followed by an analysis of the waterproof potted 20-inch PMTs and a comparison with the results from commercial electronics used by the container \#A and \#B.}


\begin{document}
\maketitle
\flushbottom

\section{Introduction}
\label{1:intro}

\begin{figure}[!htb]
    \centering
    \includegraphics[width=0.75\linewidth]{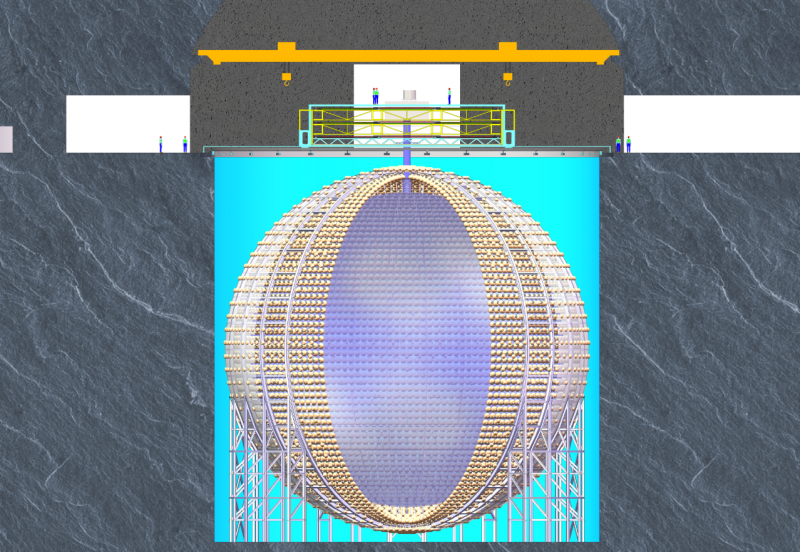}
    \caption{Schematic view of the JUNO detector\cite{2021arXiv210402565J}.}
    \label{fig:JUNO}
\end{figure}

The measurement of $\theta_{13}$ by Daya Bay\,\cite{2017PhRvD-95g2006A}, RENO\,\cite{RENO:2012mkc} and Double Chooz\,\cite{DoubleChooz:2012gmf} provides a possibility to determine the neutrino mass ordering (NMO) with reactor neutrinos. 
The Jiangmen Underground Neutrino Observatory (JUNO)\,\cite{2015arXiv150807166A,2016JPhG...43c0401A}, with a 20-kton multi-purpose underground liquid scintillator (LS) detector as shown in Fig.\ref{fig:JUNO}, plans to use 20,000 20-inch PMTs\,\cite{2019NIMPA.94762766W} (15,000 Micro Channel Plate (MCP) PMTs from Northern Night Vision Technology Ltd.\,(NNVT GDB6201)\,\cite{NNVT-GDB6201-note} and 5,000 dynode PMTs from Hamamatsu Photonics K.K. (HPK R12860)\,\cite{HPK-R12860}) for an excellent energy resolution of at least 3\% at 1\,MeV with higher than 75\% PMT coverage. The PMTs' performance has a critical impact on the measurements of the energy, timing and vertex of JUNO to ensure the NMO sensitivity.

Compared to the traditional HPK dynode PMTs, the selected 20-inch NNVT MCP PMTs were specially developed for the JUNO experiment with high quantum efficiency (QE)\cite{WANG2012113}, where the MCP, a plate-like structure consisting of many thin glass tubes with semiconductor material coated\,\cite{REN2020164333}, plays a key role for electron multiplication. There are two versions of MCP PMTs identified by their production date: the low-QE PMTs from early production and the high-QE PMTs from late production with different production process. 

Both types of JUNO selected 20-inch PMTs are required to pass a visual inspection, an acceptance test (with a pluggable HV divider\,\cite{2018arXiv180303746L,10.1007/978-981-13-1316-5_54}), and are finally potted to be water proof (HV divider firmly soldered to the PMT)\,\cite{JUNOdetector2021}. All of the 20-inch PMTs used for JUNO have been received and tested\,\cite{JUNOPMTperformance} in a container system\,\cite{2021JInst..16.8001W} and a scanning system\,\cite{2017JInst..12C6017A}. Each PMT was tested at least once during the process. A sub-sample of the potted PMTs was tested to check their features and consistency with the results of bare PMTs. 

The container system includes four containers. The container \#A and \#B are configured for the mass acceptance test with commercial electronics, by which all the qualified 20,000 20-inch PMTs of JUNO were tested at least once. The container \#D is equipped with JUNO 1F3 electronics prototype\,\cite{JUNOdetector2021} to have a combined testing with the potted 20-inch PMTs and JUNO 1F3 electronics. The 1F3 electronics is custom designed for a wide dynamic range from 1\,p.e.to 1000\,p.e. and a good linearity response\,\cite{2021arXiv211012277P}. The testing with container \#D provides early understanding of the characteristics of JUNO 1F3 electronics with potted 20-inch PMTs.

In this paper, we will present the basic features of the JUNO 1F3 electronics, and a combined characteristics of the potted 20-inch PMTs with the 1F3 electronics. A brief introduction of container \#D will be presented in Sec.\,\ref{1:system}. The analysis and results will be shown in Sec.\,\ref{1:testing}. A comparison between the results with 1F3 electronics prototype and commercial electronics is discussed in Sec.\,\ref{1:comparison}. Finally, a short summary is given in Sec.\,\ref{1:summary}. 

\section{Container \#D with 1F3 electronics prototype}
\label{1:system}

\begin{figure}[!htpb] 
\centering
\includegraphics[scale=0.7]{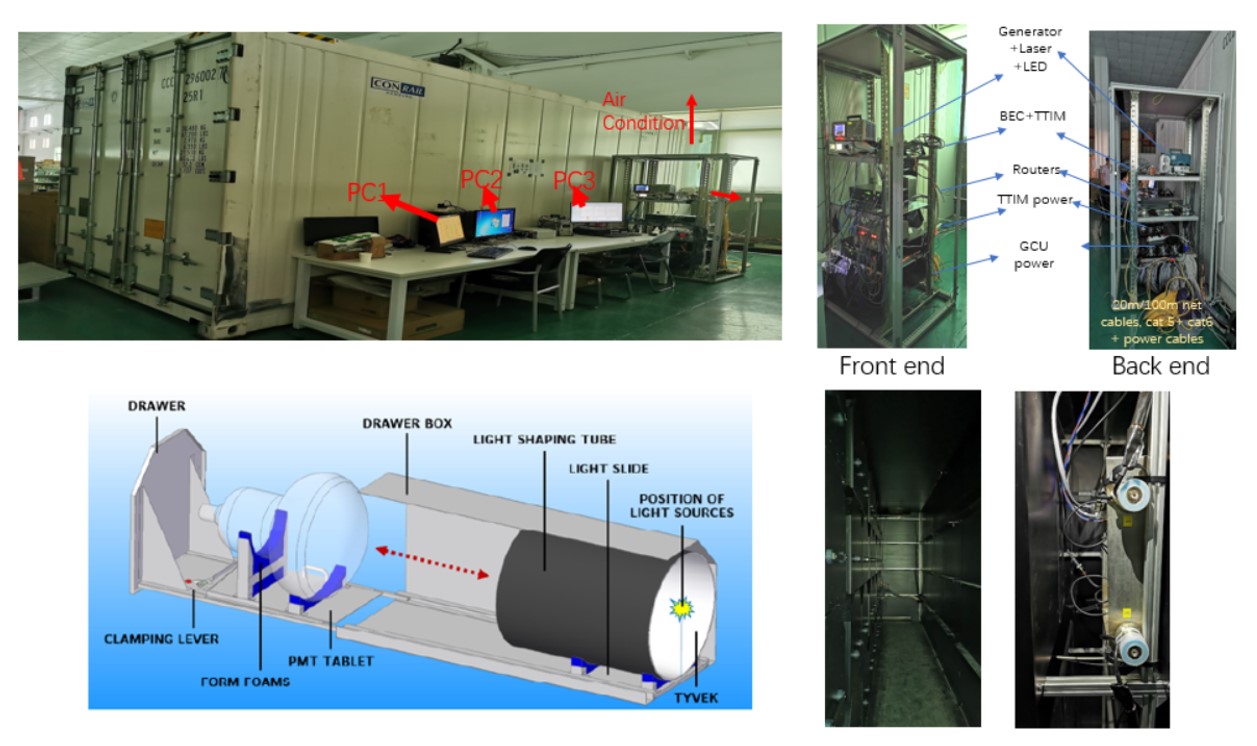}
\caption{Top left: the outer view with control PCs of the 20-feet container of container \#D. Top right: the rack where the power and net controller are installed. Bottom left: the schematic view of a drawer box. Bottom right: the internal view of the container and an installed box of JUNO 1F3 electronics between the drawers.}
\label{fig:box}
\end{figure}

The primary facility used for the qualification of the 20-inch PMTs is the container system, which has been introduced in detail in a previous paper includes the mechanical setup, data taking electronics and the measurement process\,\cite{2021JInst..16.8001W}. Only a brief description will be provided here for a better understanding on the following results, in particular for container \#D. 
The containers work as darkrooms. Each container is shielded by six alternating layers of silicon-iron to reduce the influence of the geomagnetic field, leaving the residual magnetic field only 4.7\,\textmu T which is required by the future JUNO detector too. The container \#D contains 32 testing channels (drawers) as shown in Fig.\ref{fig:box}, in which a total of 11 GCUs are installed (1 GCU for 3 electronics channels (1F3), 33 electronics channels in total). Each of the drawers is equipped with a stabilized LED light source\,\cite{HVSYS-LED}, which illuminates the full photocathode uniformly with the help of collimator, attenuator, diffuser and cylindrical reflector, and each drawer box works independently. The wavelength of the light source is 420\,nm and its intensity can be configured for different purposes\,\cite{2019NIMPA.939...61A}. A HVAC (heating, ventilation, and air conditioning) unit is used to control the measurement environment inside the containers, which is configured to 23$^\circ$C for container \#D. 

The JUNO 1F3 electronics prototype used by container \#D is integrated with a high voltage unit (HVU) and a global control unit (GCU)\footnote{An embedded version of the system can be found in \,\cite{2021NIMPA.98564600B}.} as shown in Fig.\,\ref{fig:box}. The custom HVU provides the bias voltage to PMTs. As the core of the readout electronics, the GCU is used for analog-digital conversion and waveform acquisition. 
After receiving the PMT signal, the GCU duplicates it into two streams with different amplification factors in the front end chip and converts the current signal into voltage. Then the analog signal is converted into digital signal by a 14\,bits, 1\,GS/s, custom ADC. The conversion factor from ADC to amplitude of high gain is around 0.12\,mV/ADC, which is used for the following analysis. Finally, the digital signal is further processed in the field-programmable gate array (FPGA), for local trigger generation, charge reconstruction, timestamp tagging and temporary storage.

\begin{figure}[!htpb] 
    \centering
    \includegraphics[scale=0.2]{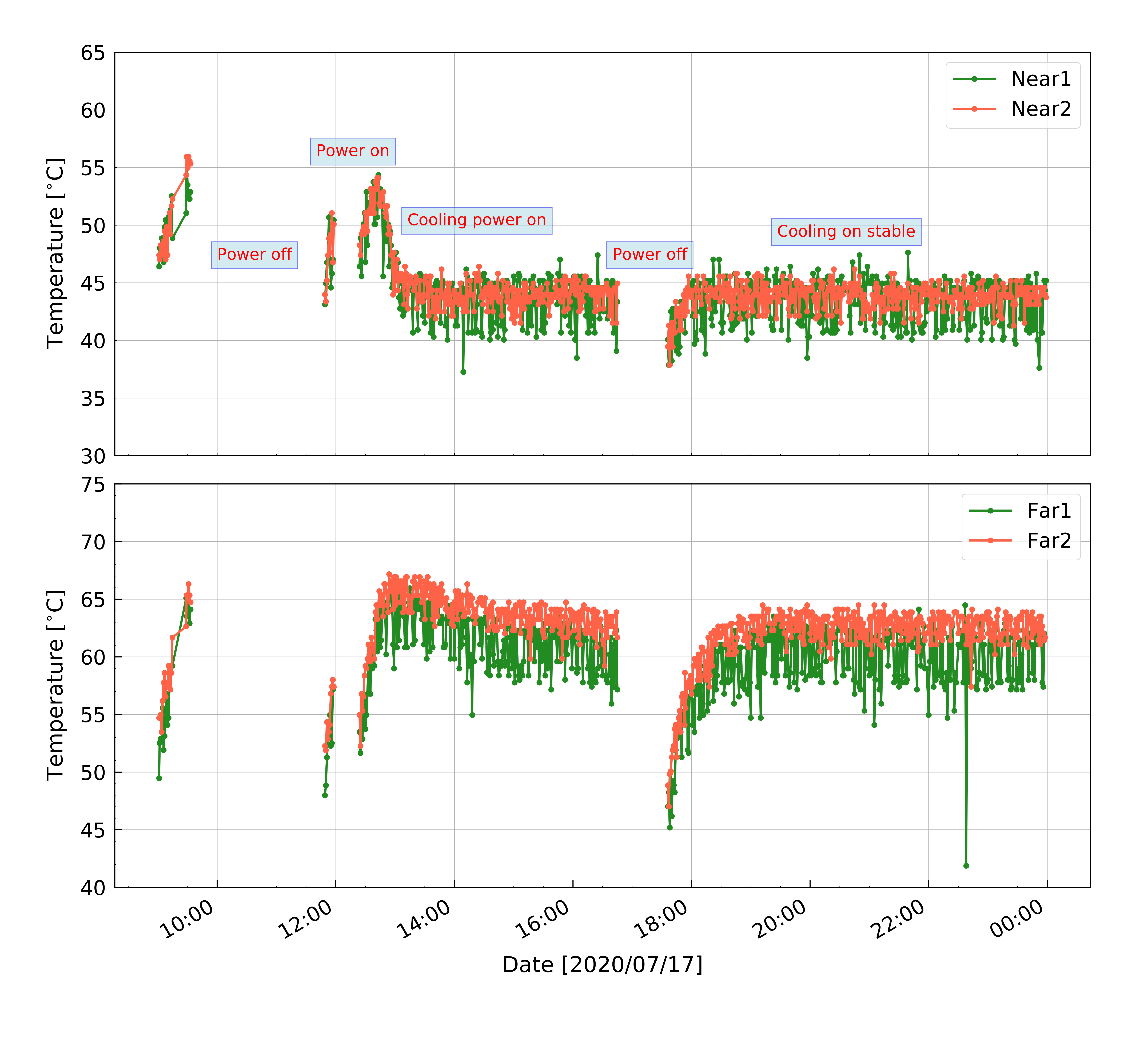}
    \caption{Temperature monitoring of the GCU channels (chips). Top: the temperature from two electronics channels from a single GCU near to the container ventilation fan; bottom: the temperature from two electronics channels from another single GCU far from the container ventilation fan.}
    \label{fig:elec:T}
\end{figure}

A DAQ system based on Linux is realized for container \#D with the 1F3 electronics\cite{junoDAQ-prototype2021}, which can provide the initialization, configurations of HVU and electronics, control and waveform readout, a temperature monitoring and a dark count rate (DCR) monitoring of each electronics channel. The temperature monitoring of each electronics channel indicates the chip temperature will reach its stable status in around 20 minutes, and can be run smoothly in the configured air environment, as shown in Fig.\,\ref{fig:elec:T}. The DCR is the pulses per second with a configurable threshold relative to the baseline.

A sub-sample of waterproof potted PMTs was tested in container \#D as well as in containers \#A or \#B following the same procedure as in \cite{JUNO-20inchPMT-testing}. A cable of 2\,m (for JUNO central detector) or 3\,m (for JUNO veto detector) with a special SHV connector is soldered to the waterproof potted 20-inch PMT of JUNO, where a stainless steel bellow is also used for pure water sealing. To make operation convenient, another 2\,m extension cable is used between the potted PMT and the 1F3 box, which however affects the measured waveform shape. 

\section{Results with container \#D}
\label{1:testing}

Various parameters of potted PMTs such as timing, amplitude and charge can be derived with the recorded waveforms. According to the testing procedure, all the parameters are checked for each PMT with container \#D, including: DCR, gain, peak-to-valley-ratio (P/V), signal-to-noise (S/N), charge resolution, amplitude, pre-pulse ratio (PPR), rise-time (RT), fall-time (FT), full width at half maximum (FWHM) and hit-time (HT) as in \cite{JUNOPMTperformance}. Please note all the potted PMT were tested with a gain of around $1\times10^7$.

\subsection{Noise level}
\label{2:noise}

\begin{figure}[!htb]
    \centering
	\begin{subfigure}[c]{0.45\textwidth}
	\centering
		\includegraphics[width=\linewidth]{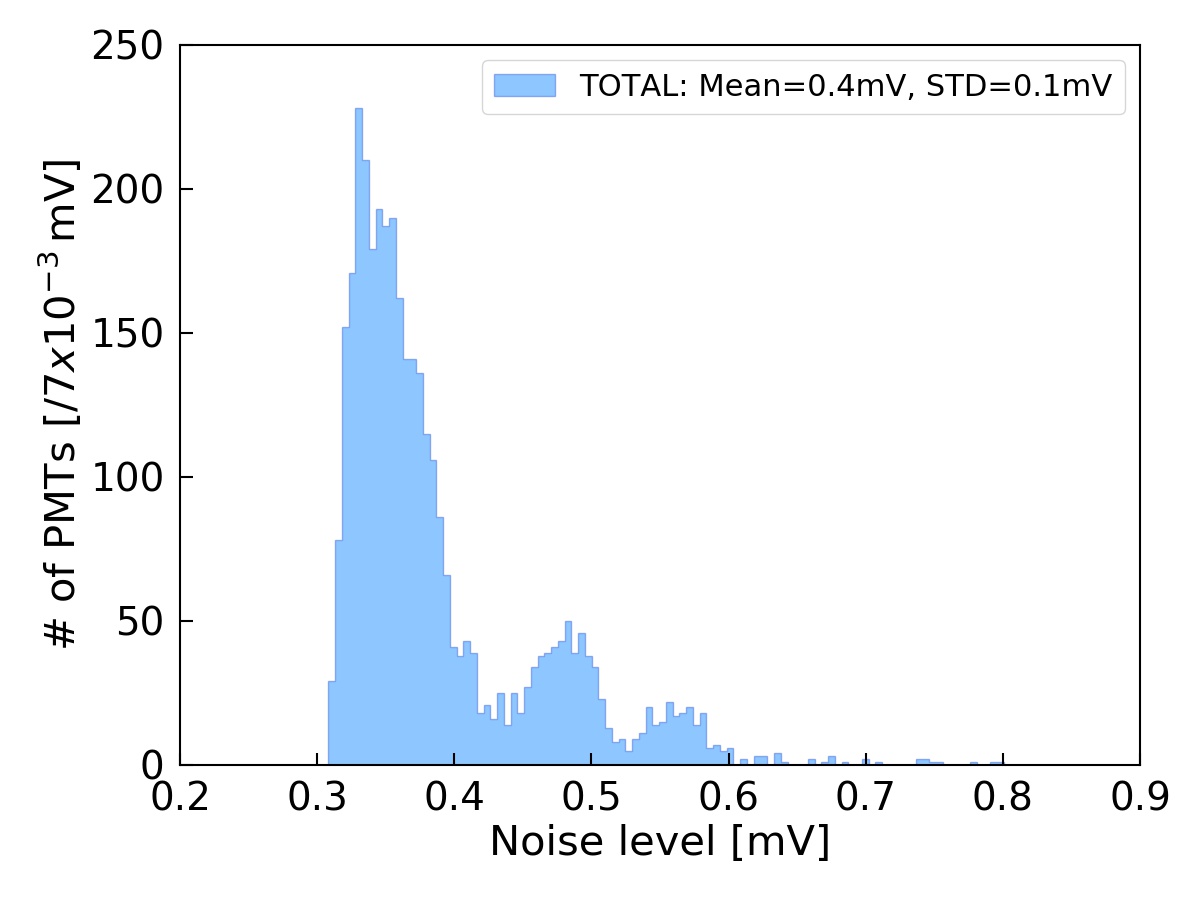}
	\end{subfigure}
	\begin{subfigure}[c]{0.45\textwidth}
	\centering
		\includegraphics[width=\linewidth]{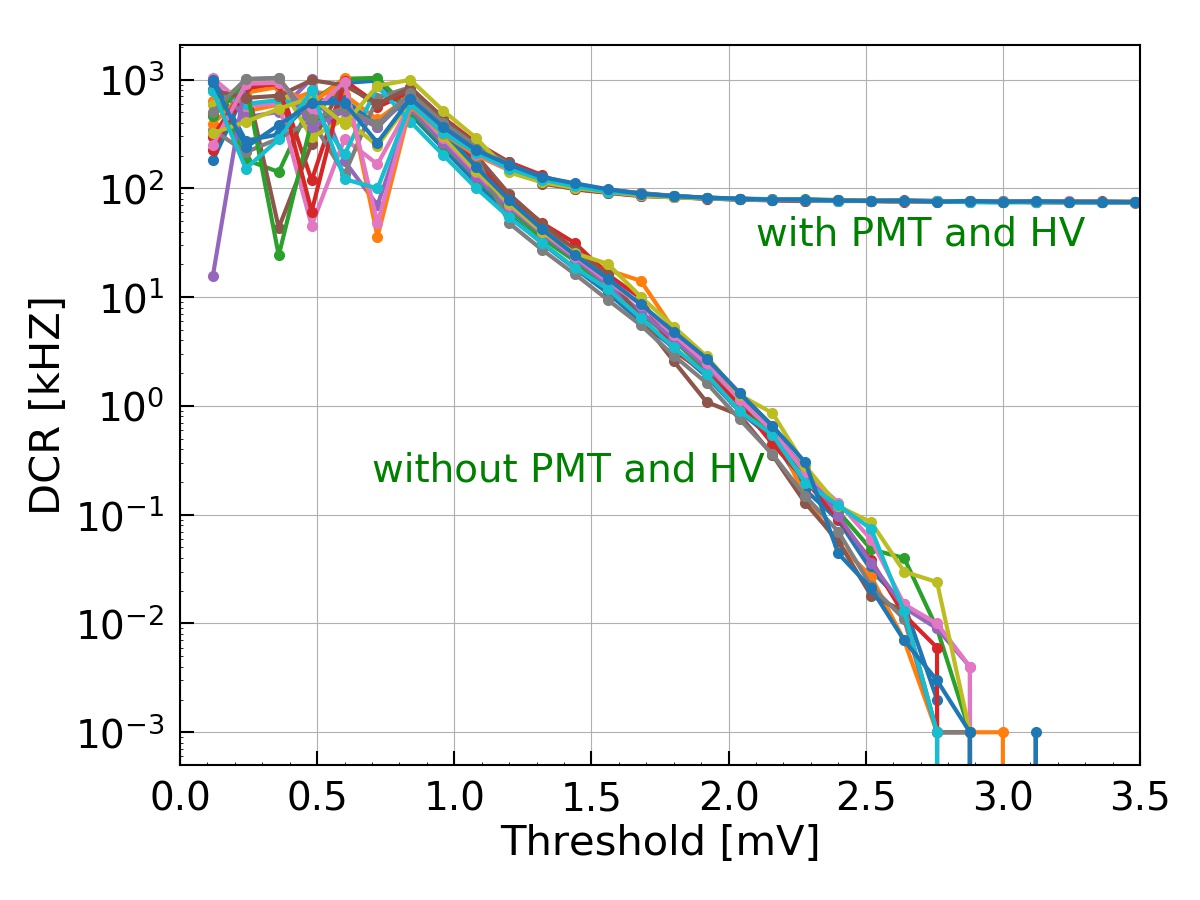}
	\end{subfigure}
    \caption{Noise level of 1F3 electronics with or without PMT connected. Left: the noise level in sigma of baseline distribution; right: the threshold survey for DCR.}
    \label{fig:noiselevel}
\end{figure}

The distribution of the waveform baseline is checked for all the 1F3 electronics channels. The sigma of a Gaussian function is used to evaluate the noise level of each electronics channel. The distribution of sigma is shown on the left of Fig.\,\ref{fig:noiselevel}. The mean noise level is around 0.4\,mV for all electronics channels without or with PMT connected and HV off/on. For some GCU boxes and PMTs, the signal baseline shows slight fluctuation, this is showing as two smaller bumps in the distribution. 

The DCR can also be used to evaluate the noise level, by checking the relationship between the counting rate and threshold. A scan of DCR vs. threshold is performed for all the electronics channels, as shown on the right of Fig.\,\ref{fig:noiselevel}. We find that the DCR decreases sharply regardless of whether PMT connected or not, after a plateau when the threshold is low\footnote{It is source from the counting saturation when the noise overs the threshold all the time.}. The threshold around 0.84\,mV indicates the edge of the noise level, which is around 2 times the noise level in sigma. A threshold around 2\,mV is suggested for PMT DCR measurement, which is around 4$\sim$5 times the noise level in sigma.

\subsection{Waveform parameters}
\label{2:par}

The digitization of PMT output wavefroms provides more opportunities for a deeper understanding on the PMT features and better precision is achieved with offline analysis. We will derive the waveform related parameters with the recorded data in single photo electron (SPE) mode such as SPE amplitude, shape and timing parameters.

\begin{figure}[!htb]
    \centering
    \includegraphics[width=0.75\linewidth]{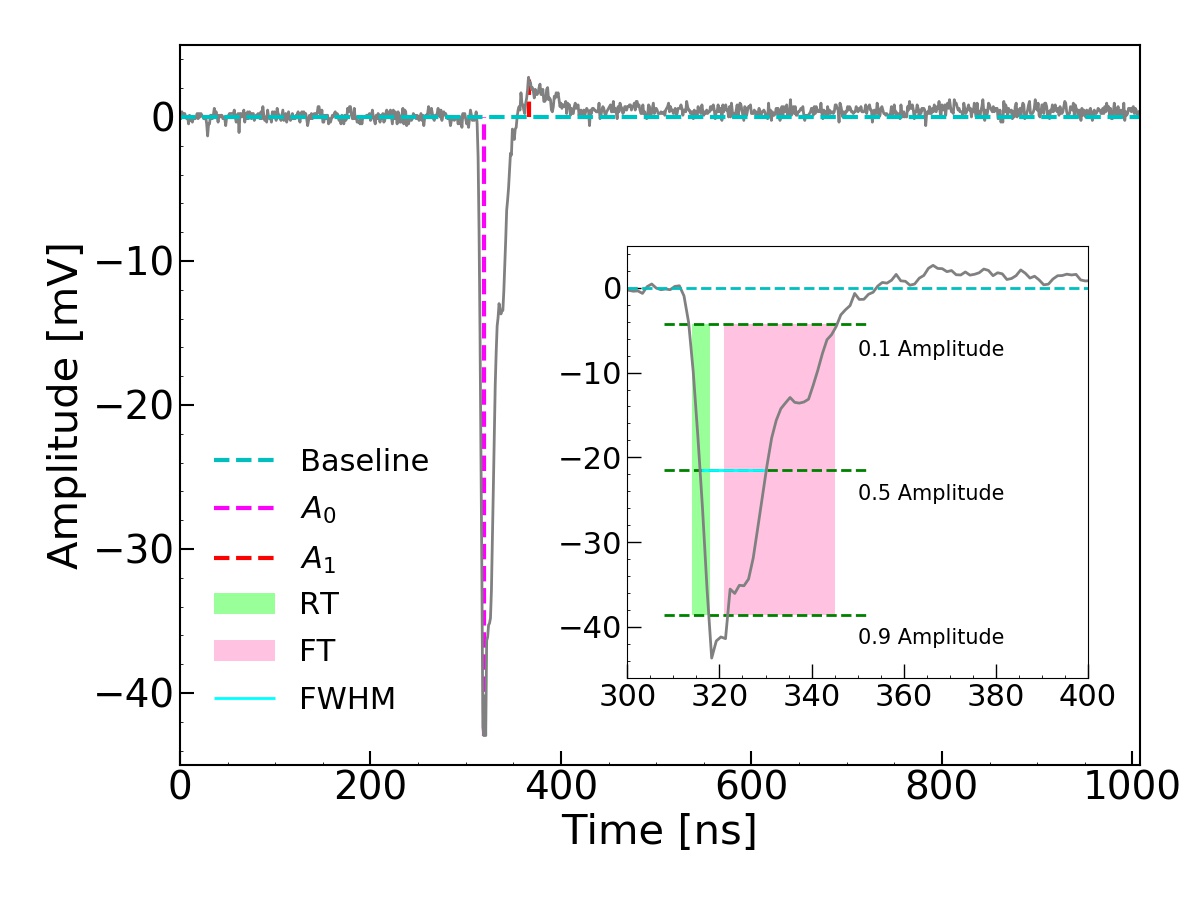}
    \caption{Exemplary of a typical NNVT waveform measured from container \#D , and related parameters labeled.}
    \label{fig:wave:overshoot}
\end{figure}

A typical waveform taken with container \#D is shown in Fig.\ref{fig:wave:overshoot}. The traditional definition of parameters is shown, where the RT is the time from 10\% to 90\% of amplitude at the rising edge of the waveform, and the FT is the time from 90\% to 10\% of amplitude at the falling edge. The FWHM is the time between 50\% of amplitude at the rising and falling edge, and the HT is the time corresponding to the threshold (3\,mV) at the rising edge. The sigma $\sigma$ of HT distribution is defined as relative TTS\footnote{There is no laser system only LED light source installed in container \#D, so the relative TTS is mainly contributed by LED system and triggering.}. An obvious amplitude overshoot can be found in the measured waveform. Few parameters are used to describe the overshoot including the primary pulse amplitude $A_0$, the peak time of the primary pulse $T_0$, the overshoot amplitude $A_1$ and its ratio to the primary pulse amplitude $A_1/A_0$, the peak time of the overshoot $T_1$ and the baseline recovery time of the overshoot $T_2$.

\begin{figure}[!htb]
    \centering
	\begin{subfigure}[c]{0.45\textwidth}
	\centering
	\includegraphics[width=\linewidth]{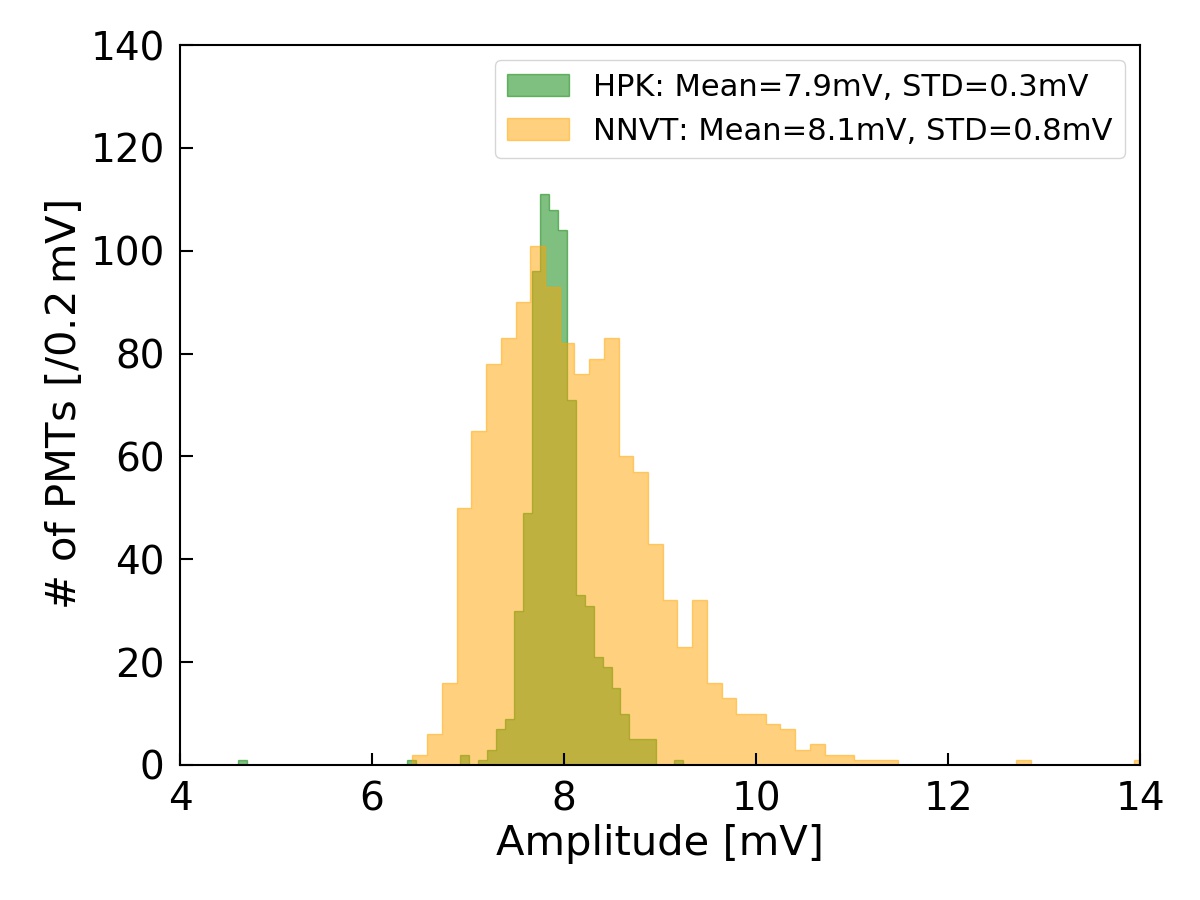}
	\caption{SPE amplitude}
	\label{fig:wave:pars:amp}
	\end{subfigure}
	\begin{subfigure}[c]{0.45\textwidth}
	\centering
	\includegraphics[width=\linewidth]{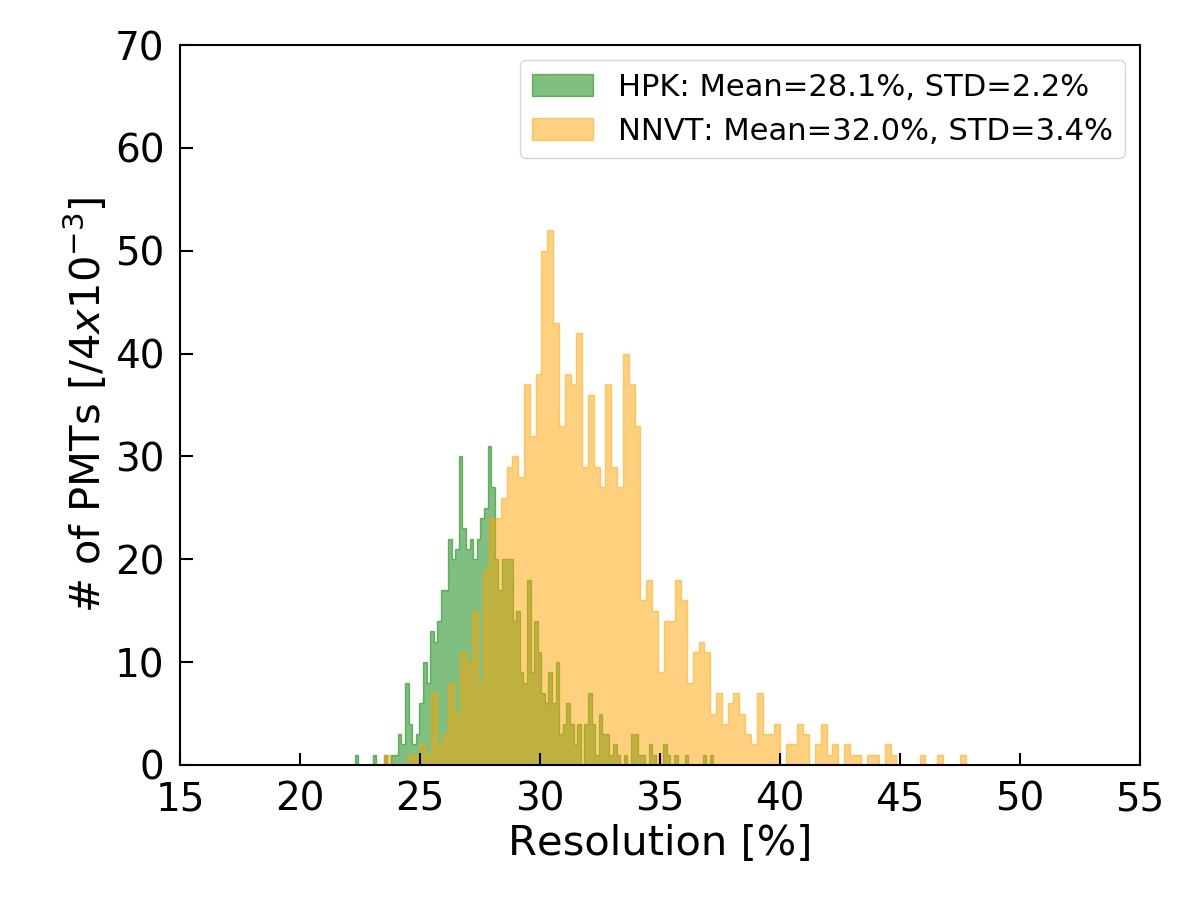}
	\caption{Resolution}
	\label{fig:wave:pars:ht}
	\end{subfigure}\\
	\centering
	\begin{subfigure}[c]{0.45\textwidth}
	\centering
	\includegraphics[width=\linewidth]{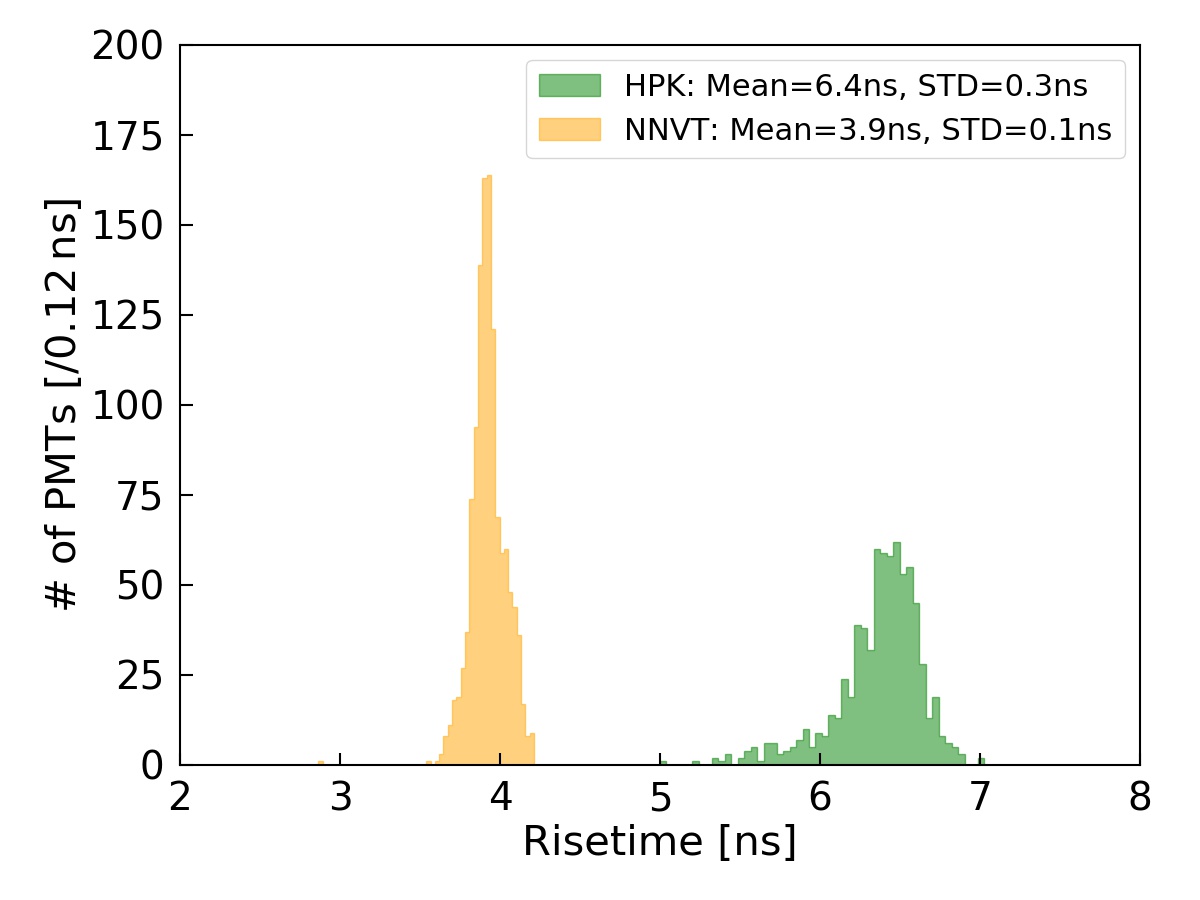}
    \caption{Rise-time}
	\label{fig:wave:pars:rt}
	\end{subfigure}
	\begin{subfigure}[c]{0.45\textwidth}
	\centering
	\includegraphics[width=\linewidth]{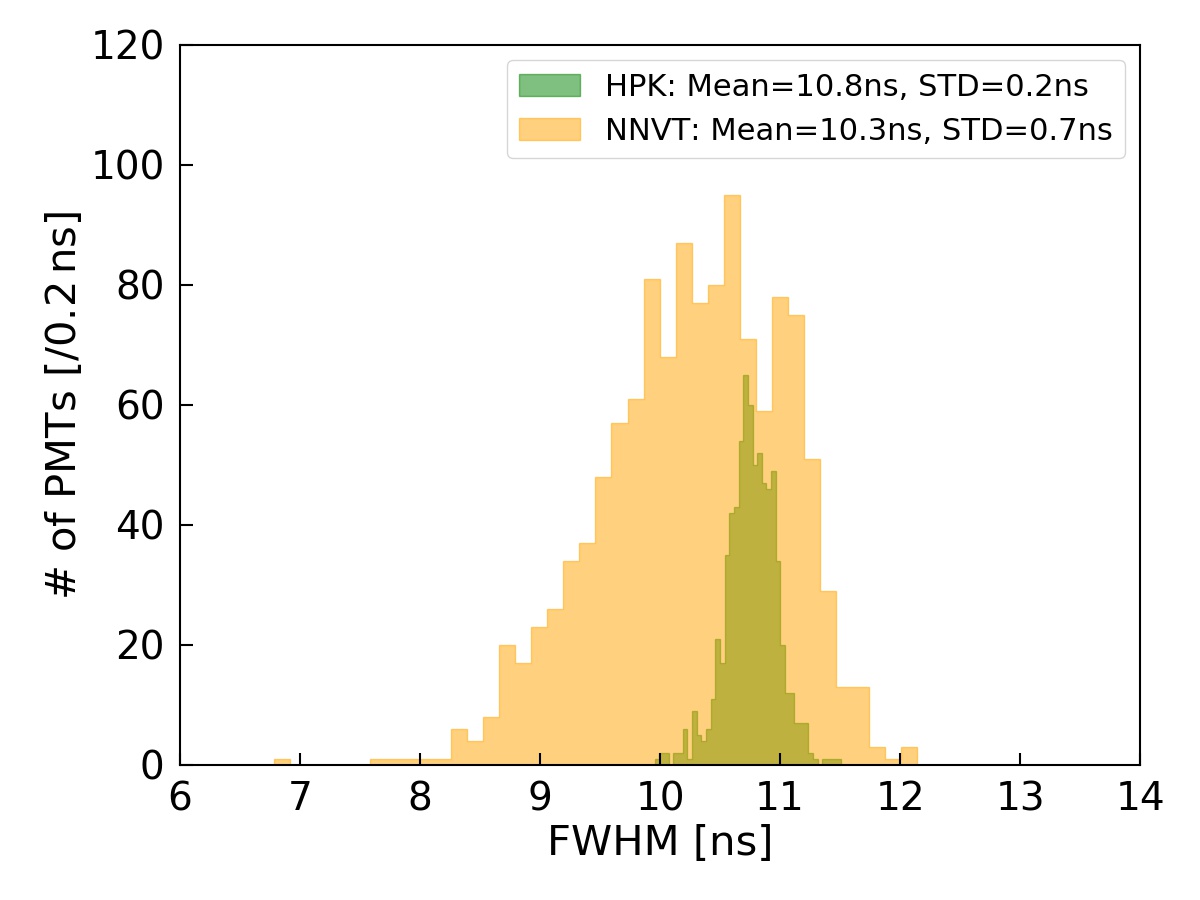}
	\caption{FWHM}
	\label{fig:wave:pars:fwhm}
	\end{subfigure}
    \caption{Waveform parameters of SPE pulses of potted PMT with 1F3 electronics. Top left: amplitude of SPE; top right: resolution; bottom left: rise-time; bottom right: FWHM. Green: HPK dynode PMTs; Orange: NNVT MCP PMTs.}
    \label{fig:wave:pars}
\end{figure}

In total 738 dynode and 1231 MCP potted PMTs were tested with container \#D, where the high-QE and low-QE MCP PMTs are 655 and 576, respectively. To derive all the listed parameters, only the waveforms with amplitude larger than 3\,mV are considered. The mean of each parameter is taken as the typical value of each PMT, some of which are shown in Fig.\,\ref{fig:wave:pars}.

\begin{figure}[!htb]
    \centering
	\begin{subfigure}[c]{0.43\textwidth}
	\centering
	\includegraphics[width=\linewidth]{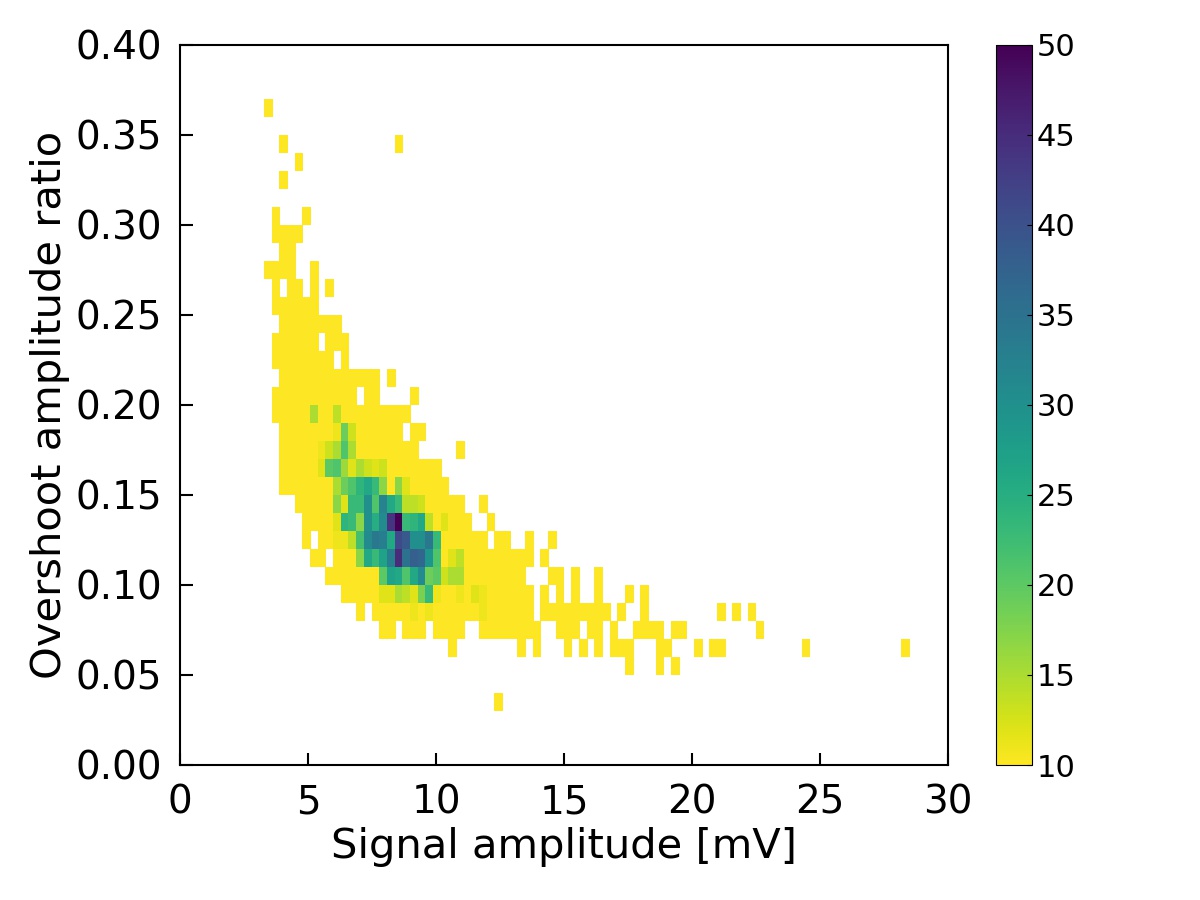}
	\caption{$A_1/A_0$ vs. $A_0$ for HPK}
	\label{fig:wave:overshoot2D:ratio:HPK}
	\end{subfigure}
	\begin{subfigure}[c]{0.43\textwidth}
	\centering
	\includegraphics[width=\linewidth]{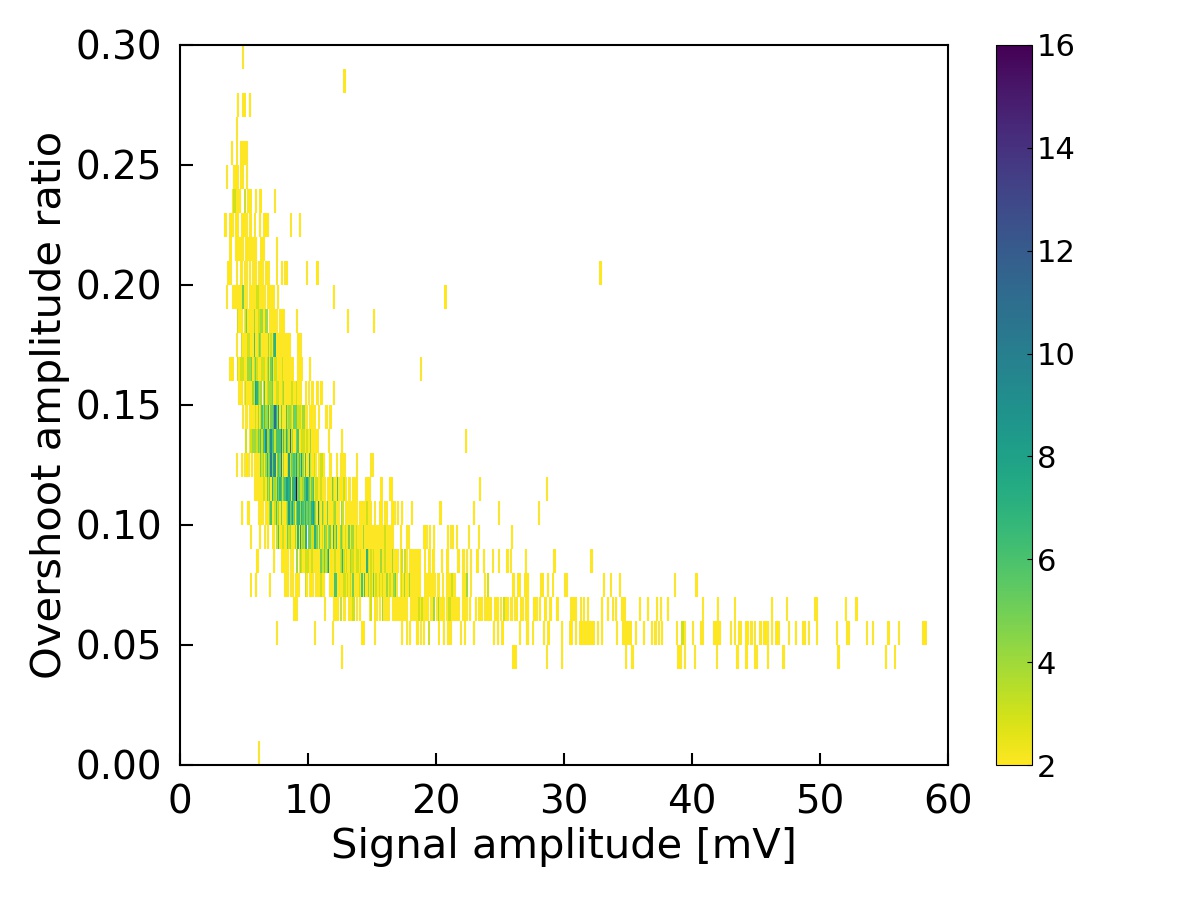}
	\caption{$A_1/A_0$ vs. $A_0$ for NNVT.}
	\label{fig:wave:overshoot2D:ratio:NNVT}
	\end{subfigure}
	\begin{subfigure}[c]{0.43\textwidth}
	\centering
	\includegraphics[width=\linewidth]{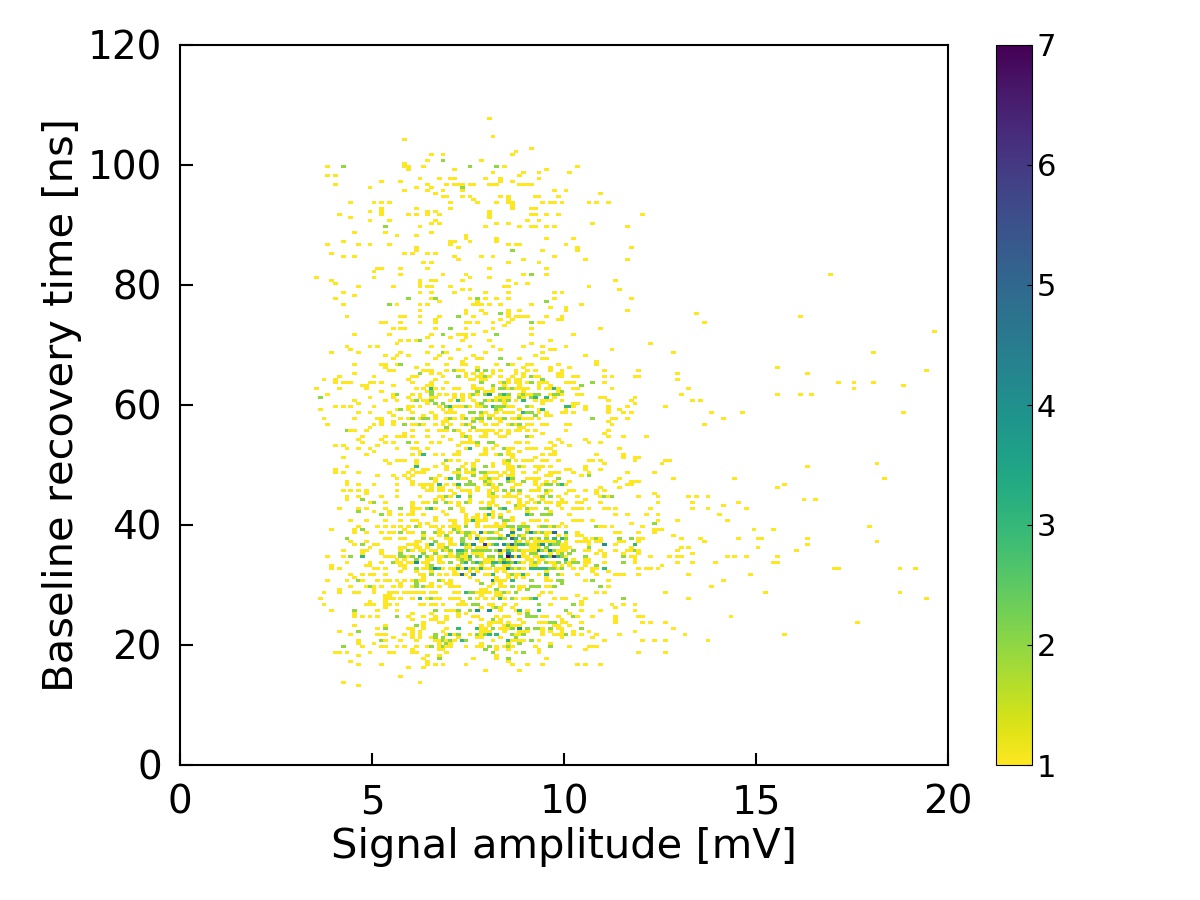}
	\caption{$T_2-T_0$ vs. $A_0$ for HPK.}
	\label{fig:wave:overshoot2D:t2:HPK}
	\end{subfigure}
	\begin{subfigure}[c]{0.43\textwidth}
	\centering
	\includegraphics[width=\linewidth]{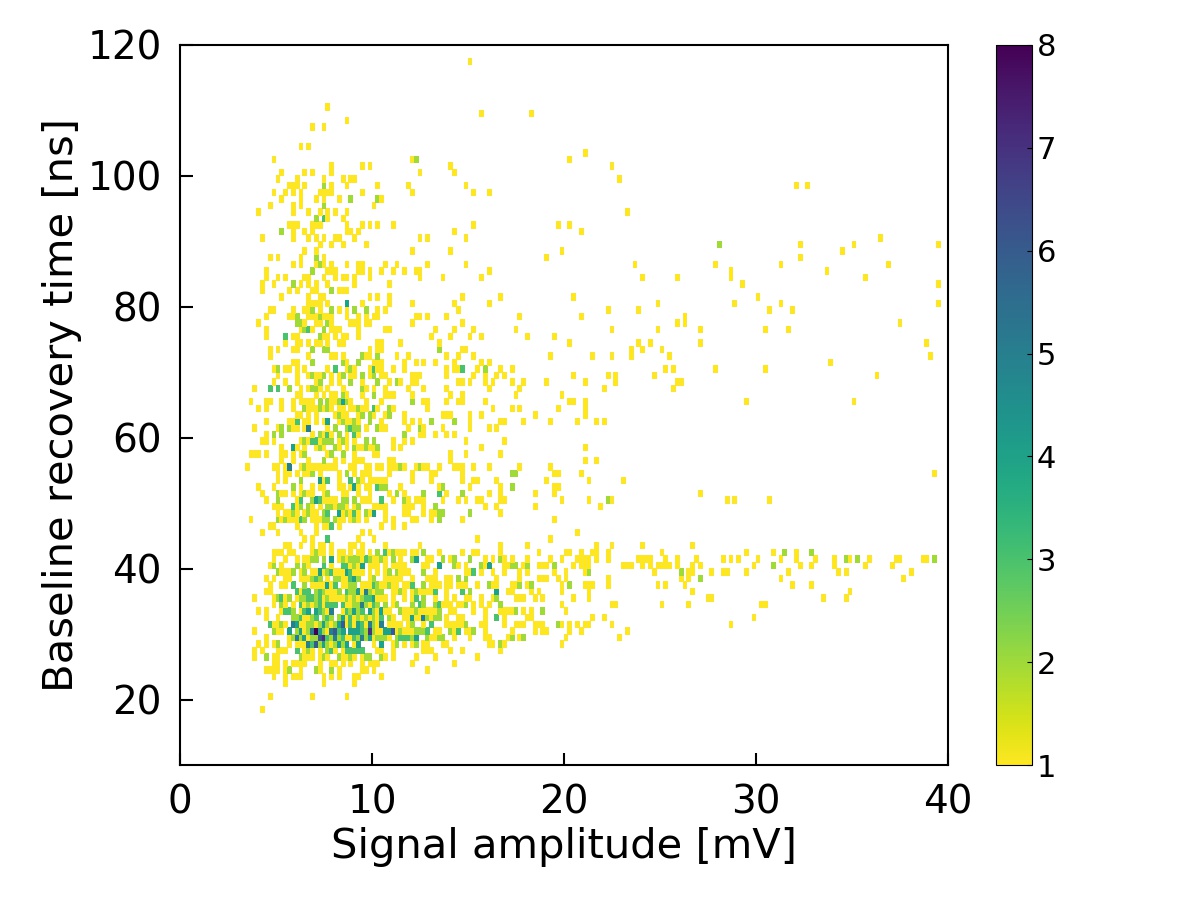}
	\caption{$T_2-T_0$ vs. $A_0$ for NNVT.}
	\label{fig:wave:overshoot2D:t2:NNVT}
	\end{subfigure}\\
    \caption{Exemplary of a single PMT on the overshoot features versus the amplitude of primary pulse. Top left: amplitude ratio of overshoot versus primary pulse amplitude for HPK; top right: amplitude ratio of overshoot versus primary pulse amplitude for NNVT; bottom left: baseline recovery time after overshoot versus primary pulse amplitude for HPK; bottom right: baseline recovery time after overshoot versus primary pulse amplitude for NNVT.}
    \label{fig:wave:overshoot2D}
\end{figure}

As known, overshoots arise from the discharging of the split capacitor for the HV and PMT pulse of the positive option of HV divider\,\cite{JUNOPMTsignalopt}. The typical overshoot values are summarised in Tab.\,\ref{tab:parameters_overshoot}, where the pulse is further selected with its waveform width larger than 5\,ns, which results in a higher mean amplitude for the sample than for the SPE. The amplitude of overshoot is convoluted with the noise, which is dominant in case of small signals as SPE. The overshoot ratio to the primary pulse in SPE is around 13\% for both types of PMTs, which is much larger than the previous testings with large signals in \cite{2016ChPhC..40i6002L,JUNOPMTsignalopt,waveAnalysisHaiqiong,JUNO-20inchPMT-testing}. An example on the correlation between the overshoot ratio versus the amplitude of the primary pulse of a single PMT is shown in Fig.\,\ref{fig:wave:overshoot2D:ratio:HPK} and \ref{fig:wave:overshoot2D:ratio:NNVT}, where an anti-correlation can be found. After an FFT\,\cite{cern-root-FFT} filtering applied on the waveforms, the overshoot amplitude ratio is reduced to around 9\% where the noise is removed partially. The overshoot is peaked at around 44\,ns after the peak of the primary pulse for both types of PMTs. The baseline of the overshoot recovers after around 49\,ns for dynode PMT and 51\,ns for MCP PMT. An example on the correlation between the recovery time versus the amplitude of the primary pulse of a single PMT is shown in Fig.\,\ref{fig:wave:overshoot2D:t2:HPK} and \ref{fig:wave:overshoot2D:t2:NNVT}, and no obvious trend is found. 

\begin{table}[!ht]
    \centering
    \caption{Typical overshoot values of the potted PMT with 1F3 electronics for the primary pulse in SPE.}
    \label{tab:parameters_overshoot}       
    \begin{tabular}{c|c|c|c|c|c}
    \hline\noalign{\smallskip}
    Types& $A_0 /mV$& $A_1 /mV$& $T_1-T_0 /ns$& $T_2-T_0 /ns$& $A_1/A_0$\\
    \noalign{\smallskip}\hline\noalign{\smallskip}
    Dynode PMT& 8.4& 1.2& 43.6& 49.0& 0.15\\
    MCP PMT& 9.9& 1.3& 43.5& 50.9& 0.12\\
    \noalign{\smallskip}\hline
    \end{tabular}
\end{table}

\begin{figure}[!htb]
    \centering
	\begin{subfigure}[c]{0.43\textwidth}
	\centering
	\includegraphics[width=\linewidth]{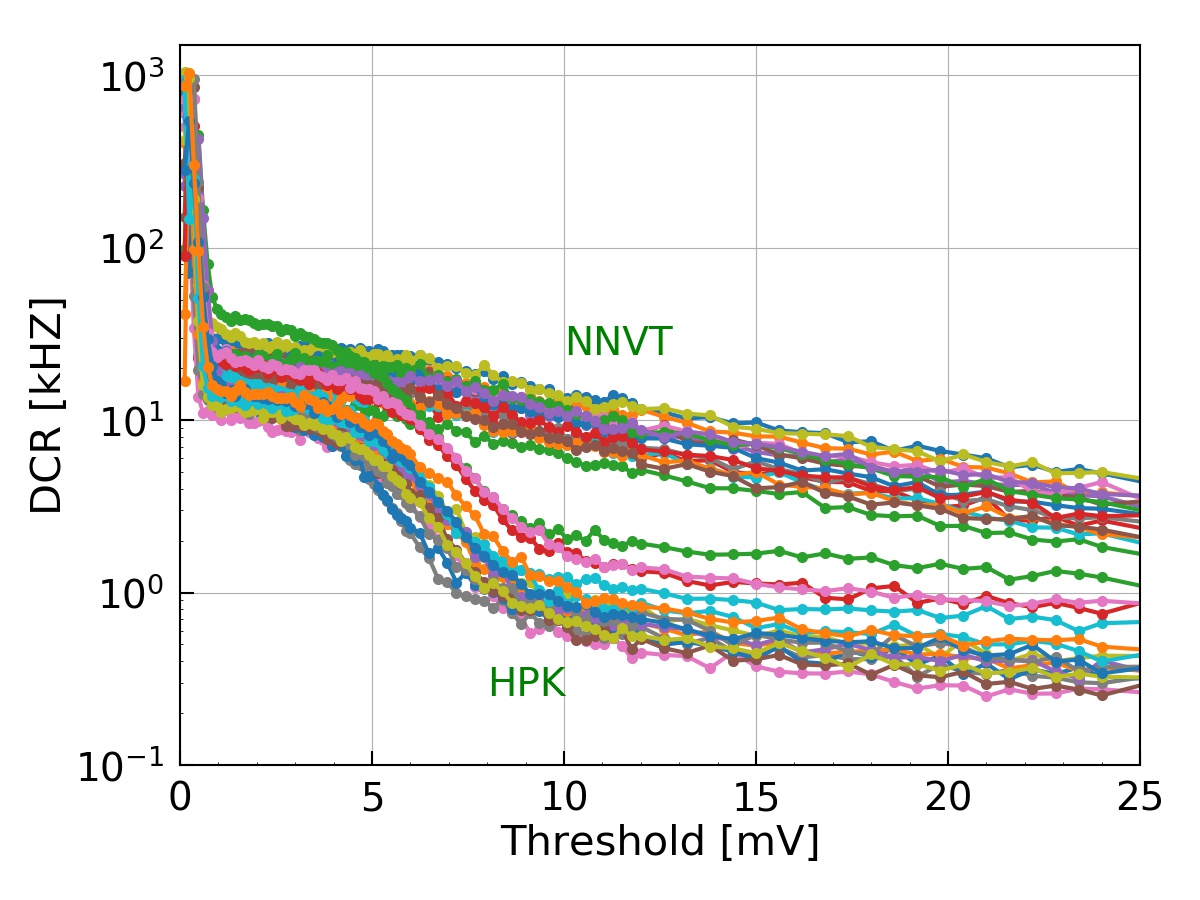}
	\caption{Threshold vs. DCR}
	\label{fig:DCR:threshold}
	\end{subfigure}
	\begin{subfigure}[c]{0.43\textwidth}
	\centering
	\includegraphics[width=\linewidth]{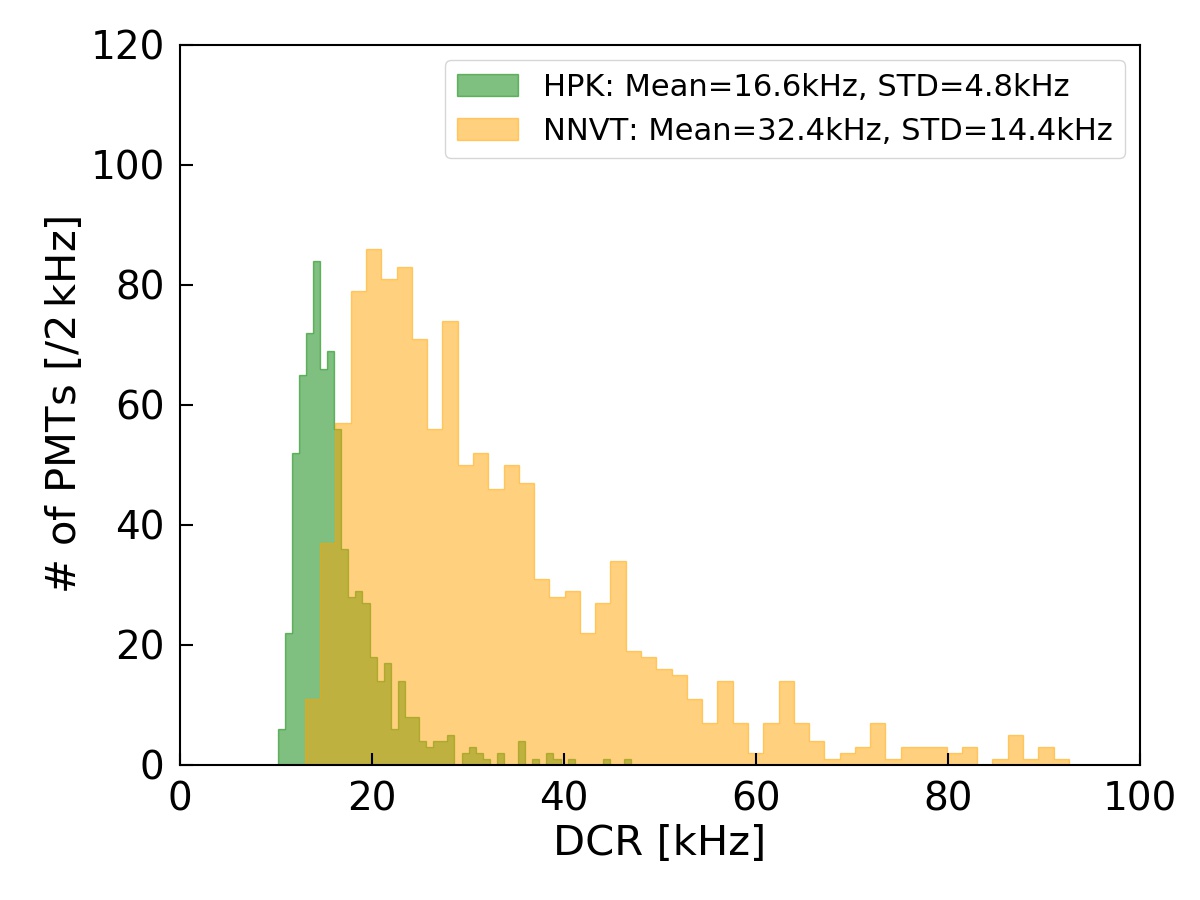}
	\caption{Measured DCR}
	\label{fig:DCR:dcr}
	\end{subfigure}
    \caption{Threshold scan for DCR and the measured DCR of the tested PMTs. Left: threshold scan with one loading of container \#D; right: measured DCR of all tested PMTs in container \#D.}
    \label{fig:DCR}
\end{figure}

The DCR of PMT is related to the applied HV, PMT's temperature and threshold. 
A threshold scan is done with one loading of container \#D (32 potted PMTs in total) as shown in Fig.\,\ref{fig:DCR:threshold}, where the NNVT MCP PMTs show a higher DCR and a systematically longer tail than HPK dynode PMT. Both types of PMTs show a large counting rate when the threshold is higher than 25\,mV, which is mainly due to events where a muon interacts with the PMT glass bulb\,\cite{zhangyu-large-pulse}. For the system noise test which is described in Sec.\,\ref{2:noise} and the SPE amplitude, an amplitude threshold of around 1.8\,mV is used for the PMT DCR measurement with container \#D. Before the DCR measurement, a cooling time of at least 12 hours is used to stabilize the DCR after the PMT loading into the container and switching on the HV, which is in the following referred to as "cooling". During the DCR testing, the calibrated HV for a gain of around $1\times10^7$ is applied. The DCR distributions for both types of PMTs are shown in Fig.\ref{fig:DCR:dcr}. The mean value is 16.6\,kHz for HPK potted PMT and 32.4\,kHz for NNVT potted PMT respectively. Please note that the temperature of container \#D is configured to 23$^{\circ}$C.

\subsection{Charge parameters}
\label{2:charge}

There are a couple of studies on the charge de-convolution of PMT waveforms \cite{dayabay-Jetter_2012,Huang-waveform,Zhang_2019_waveform,xu2021ultimate}, for simplification we use the definition from the previous analysis, the charge (Q in pC) of each waveform here is integrated offline by summing all the samples in the target window with the baseline subtracted, where the loading impedance resistance (R) of 50 Ohm and the sampling rate are considered. According to the pulse shape, a target window of 75\,ns is used: 20\,ns in front of the pulse and 55\,ns after the peak, and another 75\,ns window is used to derive the baseline which is shifted 100\,ns before the primary pulse peak.

\begin{figure}[!htb]
    \centering
	\begin{subfigure}[c]{0.43\textwidth}
	\centering
	\includegraphics[width=\linewidth]{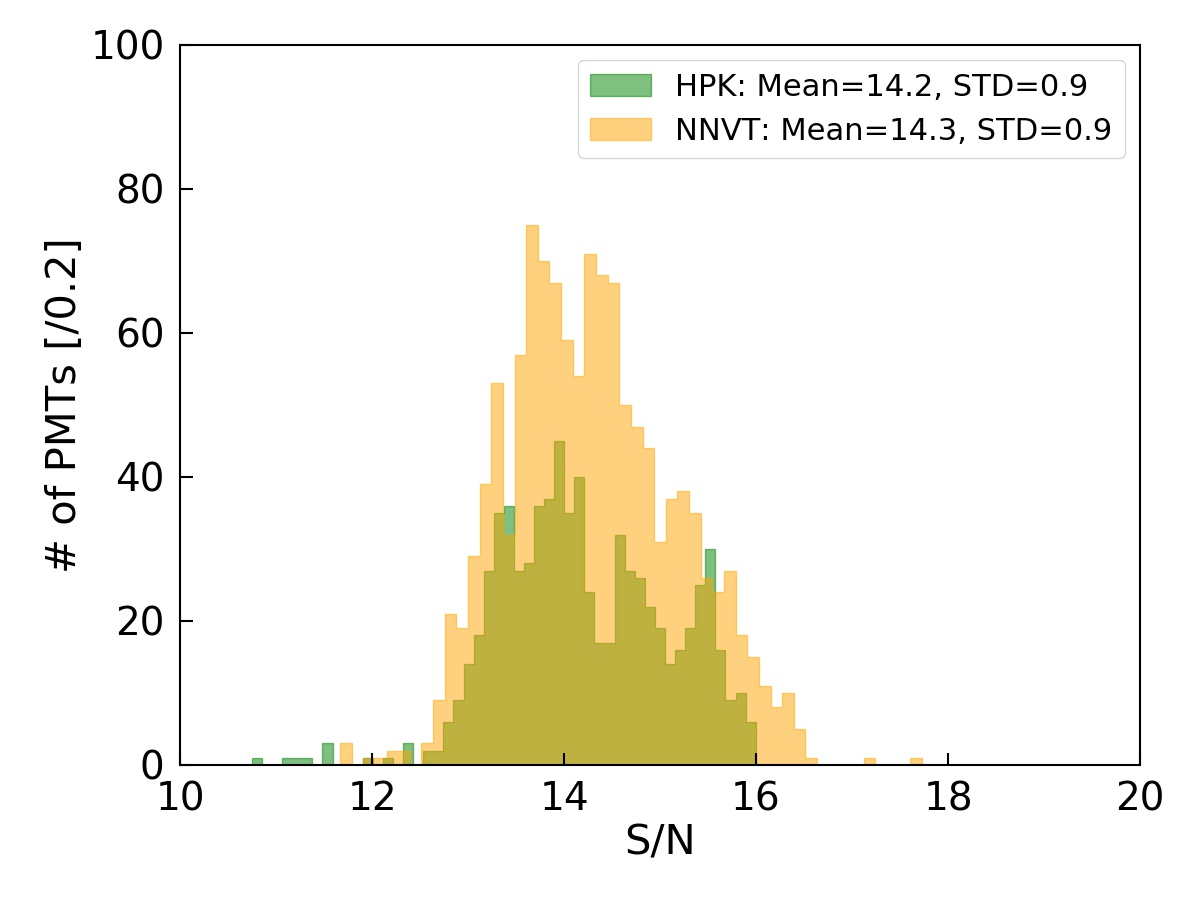}
	\caption{S/N}
	\label{fig:charge:SN}
	\end{subfigure}
	\begin{subfigure}[c]{0.43\textwidth}
	\centering
	\includegraphics[width=\linewidth]{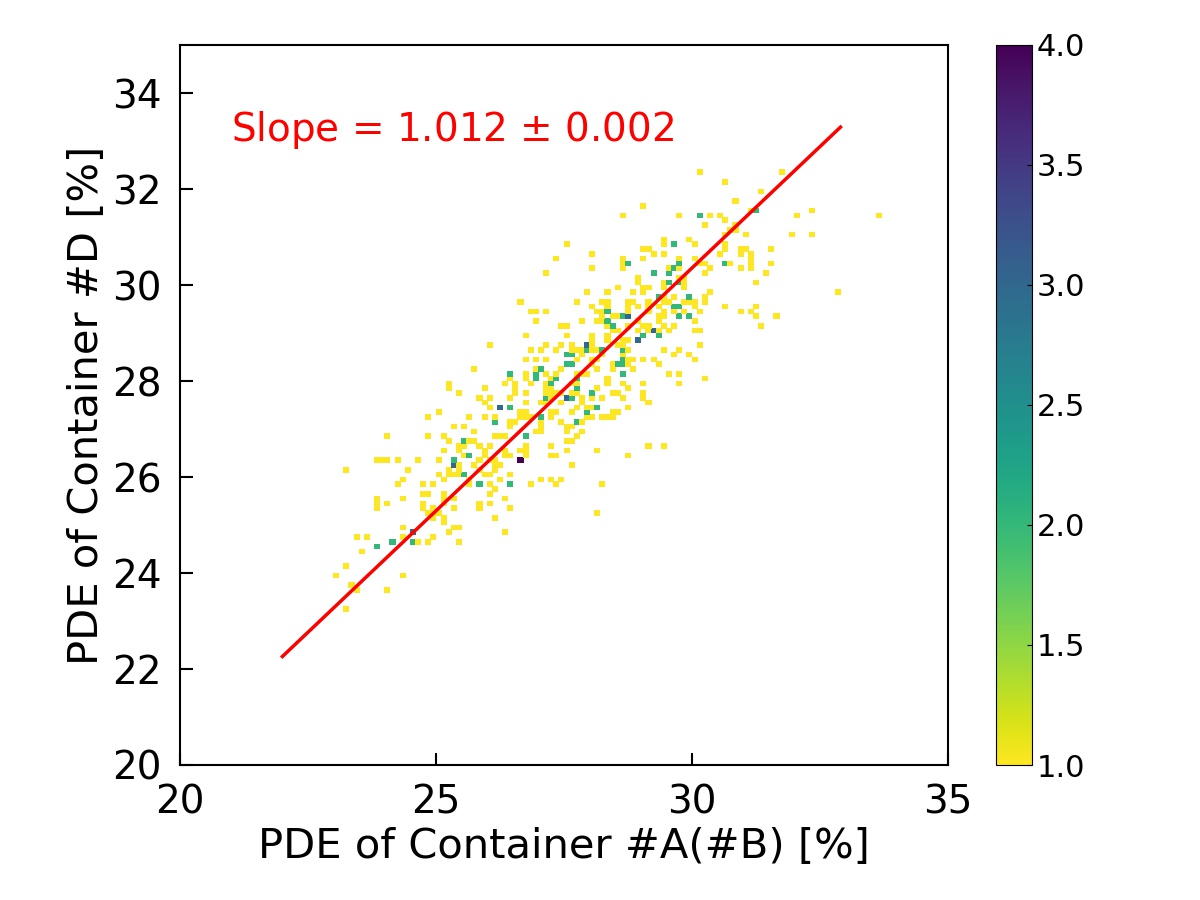}
	\caption{PDE}
	\label{fig:charge:PDE}
	\end{subfigure}
    \caption{The S/N and PDE distribution of potted PMTs in container \#D. Green: HPK
dynode PMTs; Orange: NNVT MCP PMTs.}
    \label{fig:charge}
\end{figure}

According to the testing procedure as in \cite{JUNO-20inchPMT-testing}, a working HV for a gain of around $1\times10^7$ is obtained through the HV-gain survey. The other parameters (P/V, S/N, charge resolution) were derived by SPE measurements with LED flashing following the previous definitions in \cite{JUNO-20inchPMT-testing} again. The S/N is shown in Fig.\,\ref{fig:charge:SN}.

The relative photon detection efficiency (PDE) is also determined in container \#D after calibration, where the PDE measurement is performed with the LED in multi-photon mode to minimize systematic and statistical uncertainty\,\cite{2019NIMPA.939...61A}. 
Since the measured PDE is relative, only the PDE comparison of container \#A(\#B) and \#D is shown in Fig.\ref{fig:charge:PDE}. From the fitting results, it is apparent that the two are consistent within the error range. All typical parameter values with container \#D and 1F3 electronics are collected in Tab.\,\ref{tab:parameters}. Another comparative analysis is discussed in Sec.\,\ref{1:comparison}.


\begin{table}[!ht]
    \centering
    \caption{Typical parameters of dynode and MCP PMTs in container \#D.}
    \label{tab:parameters}       
    \resizebox{\linewidth}{!}{
    \begin{tabular}{c|c|c|c|c|c}
    \hline\noalign{\smallskip}
    Parameters& All PMT& Dynode PMT& MCP PMT& High-QE PMT& Low-QE PMT\\
    \noalign{\smallskip}\hline\noalign{\smallskip}
    HV /V& 1799& 1929& 1722& 1701& 1745\\
    Gain /$10^6$& 10.0& 9.9& 10.0& 9.9& 10.1\\
    PDE /\%& 27.4& 27.7& 27.2& 29.0& 25.1\\
    DCR /kHz& 26.5& 16.6& 32.4& 31.0& 33.9\\
    Resolution /\%& 30.5& 28.0& 32.0& 32.7& 31.2\\
    P/V& 3.8& 3.6& 3.9& 3.9& 3.9\\
    FWHM /ns& 10.5& 10.8& 10.3& 10.4& 10.1\\
    S/N& 14.3& 14.2& 14.3& 14.2& 14.4\\
    RT /ns& 4.8& 6.4& 3.9& 4.0& 3.9\\
    FT /ns& 11.9& 8.9& 13.6& 14.1& 13.1\\
    HT /ns& 314.0& 285.4& 331.1& 331.8& 330.2\\
    Relative TTS /ns& 8.8& 6.2& 10.3& 10.3& 10.4\\
    Amplitude /mV& 8.1& 7.9& 8.1& 7.9& 8.4\\
    \noalign{\smallskip}\hline
    \end{tabular}
    }
\end{table}

\section{Comparison with commercial electronics}
\label{1:comparison}

\begin{figure}[!htb]
    \centering
	\begin{subfigure}[c]{0.45\textwidth}
	\centering
	\includegraphics[width=\linewidth]{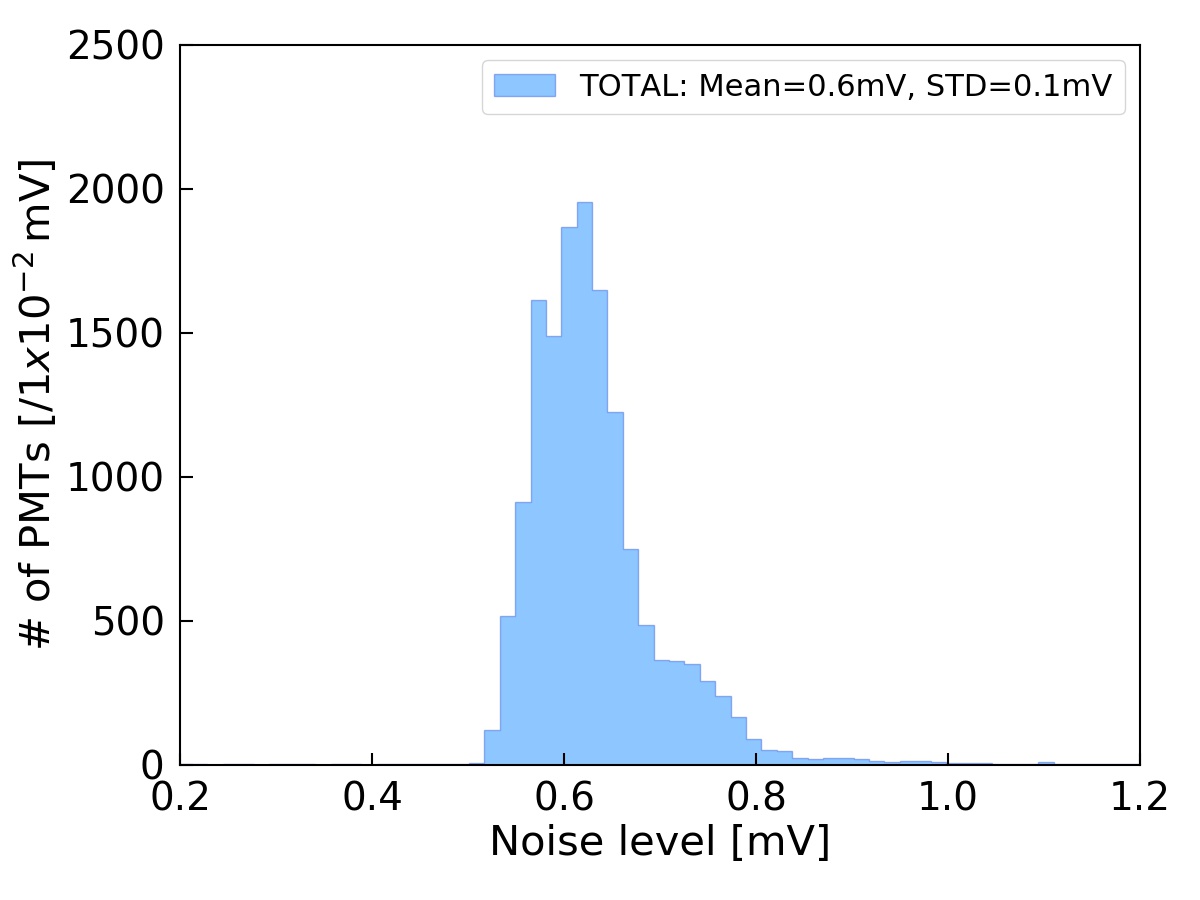}
	\end{subfigure}
	\begin{subfigure}[c]{0.45\textwidth}
	\centering
	\includegraphics[width=\linewidth]{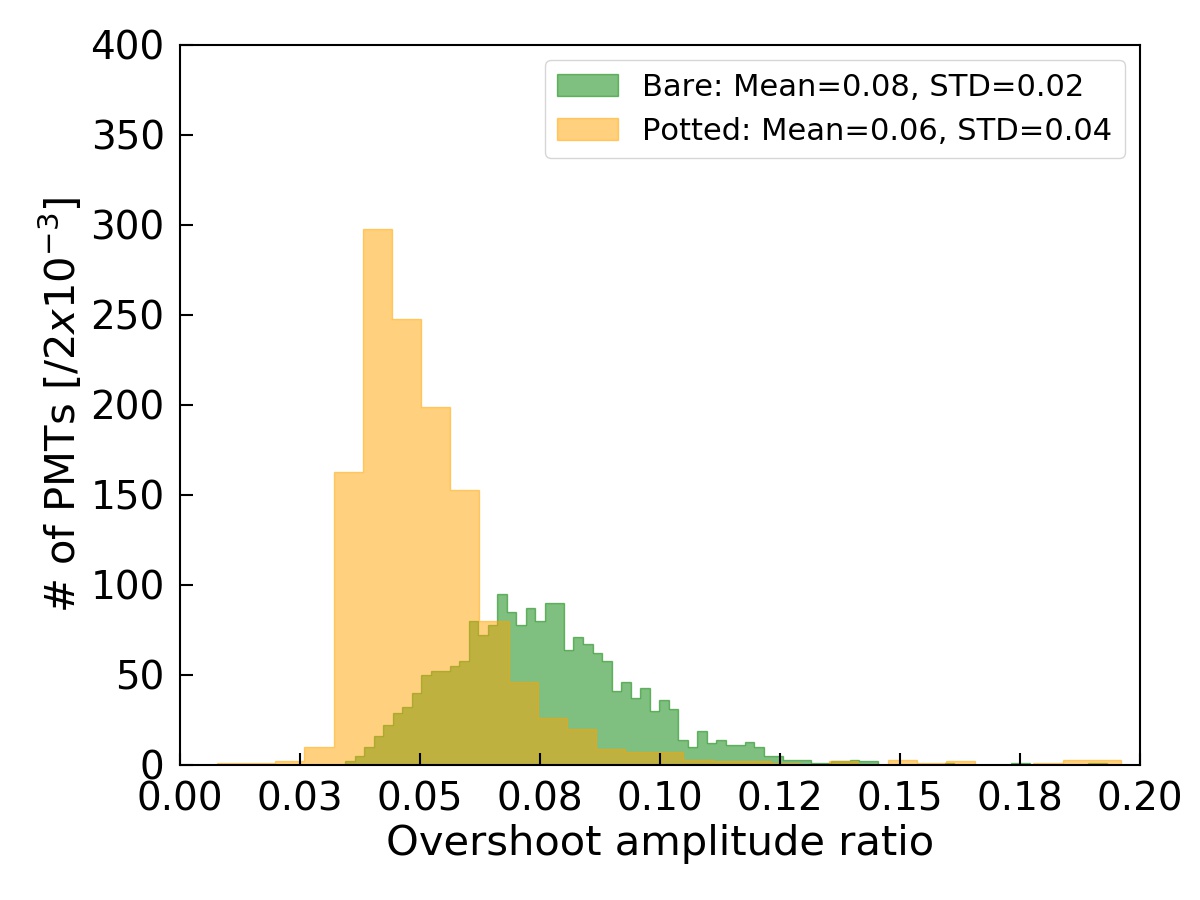}
	\end{subfigure}
    \caption{Left: noise level of baseline for containers \#A and \#B with PMT connected; right: amplitude ratio of overshoot in containers \#A and \#B with bare/potted PMTs and commercial electronics.}
    \label{fig:noiseOvershoot}
\end{figure}

\begin{figure}[!htb]
    \centering
	\includegraphics[width=0.85\linewidth]{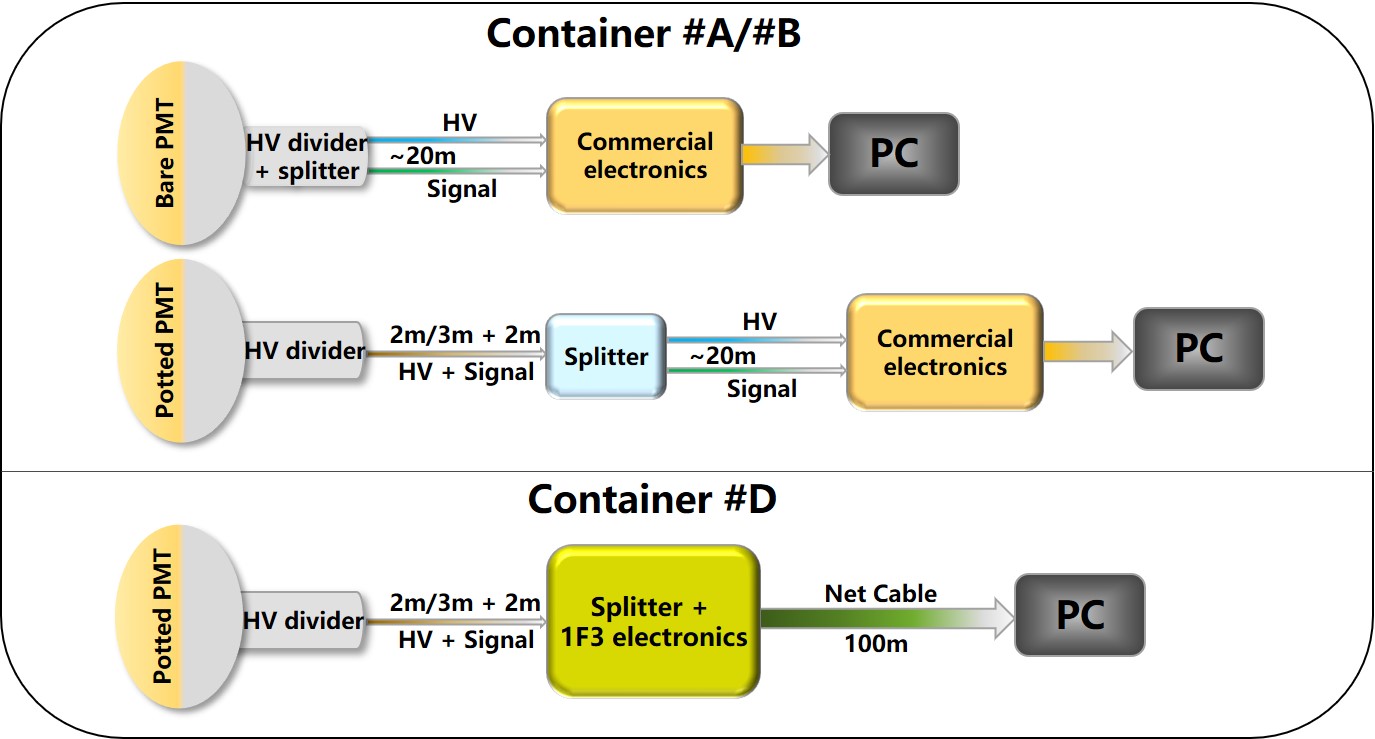}
    \caption{Comparison of cable and connections for the tests among containers \#A, \#B, and \#D for bare and potted PMTs.}
    \label{fig:electronics_com}
\end{figure}

The noise level derived from the baseline distribution of containers \#A and \#B with PMT connected is shown on the left of Fig.\,\ref{fig:noiseOvershoot}, the typical value is around 0.6\,mV, which is slightly higher than container \#D.
The amplitude ratio of overshoot for containers \#A and \#B is also checked for both bare and potted PMTs as shown on the right of Fig.\,\ref{fig:noiseOvershoot}. It should be noted that these results are average values from SPE signals, so the amplitude ratio will be larger than design due to the noise dominance. The mean value is 0.08 for bare PMTs with containers \#A and \#B and 0.06 for potted PMTs with containers \#A and \#B respectively, which is smaller than the results from container \#D with 1F3 electronics prototype and potted PMTs, in particular that of the potted PMTs. A larger overshoot will affect the charge measurement as discussed in \cite{Zhang_2019_waveform} and the baseline recovery after a huge pulse. This is mainly due to the front end cables, SHV connectors, and impedance matching inside the 1F3 electronics as shown in Fig.\,\ref{fig:electronics_com}.

To monitor the system stability, five PMTs (three dynode PMTs and two MCP PMTs, as reference PMTs) are assigned to each of the containers \#A and \#B, and one dynode PMT and one MCP PMT for container \#D. One of the reference PMTs of each container is fixed in a drawer, and the others are surveyed for all the other channels. These PMTs participated in each measurement throughout the testing task. The monitoring of the reference PMTs is used to check the reproducability and stability of the containers, which also is taken as the measurement uncertainty. Each of the parameters for every container is fitted by a Gaussian function, and its \textsigma~is taken as the test error of the corresponding container. An average value is used for container with more than one PMT of the same type. The reproducability of all parameters of dynode PMT and MCP PMT for each container is summarized in Tab.\ref{tab:referencePMT}.

\begin{table}[!ht]
    \centering
    \caption{Reproducability of dynode PMTs and MCP PMTs for containers \#A, \#B, and \#D.}
    \label{tab:referencePMT}       
    \resizebox{\linewidth}{!}{
    \begin{tabular}{c|c|c|c|c|c|c}
    \hline\noalign{\smallskip}
    \multirow{2}*{Parameters}& \multicolumn{3}{c|}{Dynode PMT} & \multicolumn{3}{c}{MCP PMT}\\
    \cline{2-7}
    \multirow{2}*{}& \#A& \#B& \#D& \#A& \#B& \#D\\
    \noalign{\smallskip}\hline\noalign{\smallskip}
    HV /V& 7.0& 6.7& 6.8& 7.1& 9.7& 15\\
    Gain /$10^6$& 0.14& 0.16& 0.11& 0.29& 0.31& 0.32\\
    PDE /\%& 0.72& 0.79& 0.47& 1.1& 0.88& 0.52\\
    DCR /kHz& 3.3& 2.0& 1.4& 7.4& 5.4& 3.3\\
    Resolution /\%& 1.7& 1.7& 1.2& 2.0& 2.2& 2.0\\
    P/V& 0.46& 0.55& 0.38& 0.58& 1.1& 0.49\\
    FWHM /ns& 0.22& 0.29& 0.25& 0.21& 0.21& 0.19\\
    S/N& 0.55& 0.36& 0.24& 0.85& 0.76& 1.1\\
    RT /ns& 0.27& 0.27& 0.11& 0.08& 0.09& 0.09\\
    FT /ns& 0.37& 0.45& 0.20& 0.55& 0.54& 0.82\\
    HT /ns& 1.5& 1.6& 2.1& 2.1& 2.5& 2.3\\
    Relative TTS /ns& 0.54& 0.50& 1.0& 1.0& 1.3& 1.0\\
    Amplitude /mV& 0.15& 0.14& 0.17& 0.34& 0.43& 0.23\\
    \noalign{\smallskip}\hline
    \end{tabular}
    }
\end{table}

To further quantify the difference, a parameter $R$ is defined as Eq.\ref{eq:R} to describe the relative inconsistency, where $P_D$ and $P_{A(B)}$ are the values of PMTs tested in containers \#D and \#A (or \#B) respectively, and $\mathop{P}\limits^{-} = \frac{1}{2}(P_D + P_{A(B)})$ is the mean. The $R$ of each parameter is obtained by taking the mean value after filling the $R$ of all PMTs into a histogram. And the error of $R$ can be calculated by Eq.\ref{eq:R_error}, which combines the measurement error of containers and the error transfer formula. The $R$ and its error of the two types of PMTs are summarized in Tab.\ref{tab:Randerror}. Ignoring HT, the R values of all parameters are almost zero within the range of [-$\Delta_{R}$, +$\Delta_{R}$], even most $R$ values are accurate to the second and third decimal number. Therefore, we deduce that the parameter responses of the dynode PMT and MCP PMT with the commercial electronics and 1F3 electronics prototype are consistent within the range of [-$\Delta_{R}$, +$\Delta_{R}$].

\begin{equation}
R = \frac{P_D - \mathop{P}\limits^{-}}{\mathop{P}\limits^{-}}   
\label{eq:R}
\end{equation}

\begin{equation}
\Delta_{R} = \frac{\sqrt{{(P_{A(B)}\sigma_{P_D})}^2 + {({P_D}\sigma_{P_{A(B)}})}^2}}{2{\mathop{P}\limits^{-}}^2}     
\label{eq:R_error}
\end{equation}

\begin{table}[!ht]
    \centering
    \caption{The R and error of dynode PMT and MCP PMT.}
    \label{tab:Randerror}       
    \resizebox{\linewidth}{!}{
    \begin{tabular}{c|c|c}
    \hline\noalign{\smallskip}
    Parameters& Dynode PMT& MCP PMT\\
    \noalign{\smallskip}\hline\noalign{\smallskip}
    HV& 0.007 $\pm$ 0.003& 0.002 $\pm$ 0.005\\
    Gain& -0.002 ± 0.009& -0.009 ± 0.022\\
    PDE& 0.006 $\pm$ 0.016& 0.008 $\pm$ 0.020\\
    DCR& 0.03 $\pm$ 0.11& -0.04 $\pm$ 0.13\\
    Resolution& -0.005 $\pm$ 0.037& -0.007 $\pm$ 0.045\\
    P/V& -0.03 $\pm$ 0.09& -0.02 $\pm$ 0.13\\
    FWHM& 0.05 $\pm$ 0.02& 0.04 $\pm$ 0.01\\
    S/N& 0.05 $\pm$ 0.02& 0.04 $\pm$ 0.05\\
    RT& 0.004 $\pm$ 0.023& 0.05 $\pm$ 0.17\\
    FT& -0.05 $\pm$ 0.02& -0.05 $\pm$ 0.04\\
    HT& 0.21 $\pm$ 0.005& 0.190 $\pm$ 0.006\\
    Relative TTS& -0.006 $\pm$ 0.091& 0.02 $\pm$ 0.08\\
    Amplitude& 0.10 $\pm$ 0.02& 0.07 $\pm$ 0.03\\
    \noalign{\smallskip}\hline
    \end{tabular}
    }
\end{table}

\section{Summary}
\label{1:summary}

With this study, the basic features of JUNO 1F3 electronics prototype used for 20-inch PMT testing of container \#D at JUNO Pan-Asia 20-inch PMT testing station are checked, including noise level and PMT parameters. Among the PMT parameters calculated with the container \#D 1F3 electronics, most of them except the overshoot ratio are consistent with previous acceptance tests with commercial electronics. The overshoot ratio in case of SPE signal is enlarged than original design due to the extension cables, connectors and impedance matching inside the electronics. For a relative large overshoot, it will affect the baseline recovery time after large signals, and affect the charge calculation caused non-linearity, which still needs further validation in detail.



\section*{Acknowledgments}
\label{sec:1:acknow}

This work was supported by the National Natural Science Foundation (NSFC) of China No. 11875282, the Strategic Priority Research Program of the Chinese Academy of Sciences, Grant No. XDA10011100, the CAS Center for Excellence in Particle Physics, the Joint Institute of Nuclear Research (JINR), Russia and Lomonosov Moscow State University in Russia, the joint Russian Science Foundation (RSF), DFG (Deutsche Forschungsgemeinschaft).
The authors acknowledge all colleagues from JUNO collaboration for operating the 20-inch PMT testing system.

\bibliographystyle{unsrtnat}
\bibliography{ref}   

\begin{thebibliography}{32}
\providecommand{\natexlab}[1]{#1}
\providecommand{\url}[1]{\texttt{#1}}
\expandafter\ifx\csname urlstyle\endcsname\relax
  \providecommand{\doi}[1]{doi: #1}\else
  \providecommand{\doi}{doi: \begingroup \urlstyle{rm}\Url}\fi

\bibitem[{An} et~al.(2017)]{2017PhRvD-95g2006A}
F.~P. {An} et~al.
\newblock {Measurement of electron antineutrino oscillation based on 1230 days
  of operation of the Daya Bay experiment}.
\newblock \emph{Phys. Rev. D}, 95\penalty0 (7):\penalty0 072006, April 2017.
\newblock \doi{10.1103/PhysRevD.95.072006}.

\bibitem[Ahn et~al.(2012)]{RENO:2012mkc}
J.~K. Ahn et~al.
\newblock {Observation of Reactor Electron Antineutrino Disappearance in the
  RENO Experiment}.
\newblock \emph{Phys. Rev. Lett.}, 108:\penalty0 191802, 2012.
\newblock \doi{10.1103/PhysRevLett.108.191802}.

\bibitem[Abe et~al.(2012)]{DoubleChooz:2012gmf}
Y.~Abe et~al.
\newblock {Reactor electron antineutrino disappearance in the Double Chooz
  experiment}.
\newblock \emph{Phys. Rev. D}, 86:\penalty0 052008, 2012.
\newblock \doi{10.1103/PhysRevD.86.052008}.

\bibitem[{Adam} et~al.(2015)]{2015arXiv150807166A}
T.~{Adam} et~al.
\newblock {JUNO Conceptual Design Report}.
\newblock \emph{arXiv e-prints}, art. arXiv:1508.07166, August 2015.

\bibitem[{An} et~al.(2016)]{2016JPhG...43c0401A}
Fengpeng {An} et~al.
\newblock {Neutrino physics with JUNO}.
\newblock \emph{Journal of Physics G Nuclear Physics}, 43\penalty0
  (3):\penalty0 030401, March 2016.
\newblock \doi{10.1088/0954-3899/43/3/030401}.

\bibitem[{Wen} et~al.(2019)]{2019NIMPA.94762766W}
L.~J. {Wen} et~al.
\newblock {A quantitative approach to select PMTs for large detectors}.
\newblock \emph{Nuclear Instruments and Methods in Physics Research A},
  947:\penalty0 162766, December 2019.
\newblock \doi{10.1016/j.nima.2019.162766}.

\bibitem[Ltd.(2019)]{NNVT-GDB6201-note}
Northern Night Vision~Technology Ltd.
\newblock {Specification for GDB-6201 microchannel plate type photomultiplier
  PMT (in Chinese)}, 2019.
\newblock https://max.book118.com/html/2020/0214/7002125152002115.shtm.

\bibitem[K.K.(2019)]{HPK-R12860}
Hamamatsu~Photonics K.K.
\newblock {R12860 datasheet}, 2019.
\newblock https://www.hamamatsu.com/jp/en/product/type/R12860/index.html.

\bibitem[{JUNO Collaboration} et~al.(2021)]{2021arXiv210402565J}
{JUNO Collaboration} et~al.
\newblock {JUNO Physics and Detector}.
\newblock \emph{arXiv e-prints}, art. arXiv:2104.02565, April 2021.

\bibitem[Wang et~al.(2012)]{WANG2012113}
Yifang Wang et~al.
\newblock A new design of large area mcp-pmt for the next generation neutrino
  experiment.
\newblock \emph{Nuclear Instruments and Methods in Physics Research Section A:
  Accelerators, Spectrometers, Detectors and Associated Equipment},
  695:\penalty0 113--117, 2012.
\newblock ISSN 0168-9002.
\newblock \doi{https://doi.org/10.1016/j.nima.2011.12.085}.
\newblock URL
  \url{https://www.sciencedirect.com/science/article/pii/S0168900211023199}.
\newblock New Developments in Photodetection NDIP11.

\bibitem[Ren et~al.(2020)]{REN2020164333}
Ling Ren et~al.
\newblock Study on the improvement of the 20-inch microchannel plate
  photomultiplier tubes for neutrino detector.
\newblock \emph{Nuclear Instruments and Methods in Physics Research Section A:
  Accelerators, Spectrometers, Detectors and Associated Equipment},
  977:\penalty0 164333, 2020.
\newblock ISSN 0168-9002.
\newblock \doi{https://doi.org/10.1016/j.nima.2020.164333}.
\newblock URL
  \url{https://www.sciencedirect.com/science/article/pii/S0168900220307300}.

\bibitem[{Luo} et~al.(2018){Luo}, {Wang}, {Qin}, and
  {Heng}]{2018arXiv180303746L}
Fengjiao {Luo}, Zhimin {Wang}, Zhonghua {Qin}, and Yuekun {Heng}.
\newblock {Signal Optimization with HV divider of MCP-PMT for JUNO}.
\newblock \emph{arXiv e-prints}, art. arXiv:1803.03746, March 2018.

\bibitem[Qin(2018)]{10.1007/978-981-13-1316-5_54}
Zhonghua Qin.
\newblock Status of the 20-in. pmt instrumentation for the juno experiment.
\newblock In Zhen-An Liu, editor, \emph{Proceedings of International Conference
  on Technology and Instrumentation in Particle Physics 2017}, pages 285--293,
  Singapore, 2018. Springer Singapore.

\bibitem[Collaboration(2021)]{JUNOdetector2021}
JUNO Collaboration.
\newblock Juno physics and detector.
\newblock \emph{Progress in Particle and Nuclear Physics}, page 103927, 2021.
\newblock ISSN 0146-6410.
\newblock \doi{https://doi.org/10.1016/j.ppnp.2021.103927}.
\newblock URL
  \url{https://www.sciencedirect.com/science/article/pii/S0146641021000880}.

\bibitem[collaboration(2022)]{JUNOPMTperformance}
JUNO collaboration.
\newblock Mass testing and characterization of 20-inch {PMTs} for {JUNO}, 2022.
\newblock https://doi.org/10.48550/arXiv.2205.08629.

\bibitem[{Wonsak} et~al.(2021{\natexlab{a}})]{2021JInst..16.8001W}
B.~{Wonsak} et~al.
\newblock {A container-based facility for testing 20'000 20-inch PMTs for
  JUNO}.
\newblock \emph{Journal of Instrumentation}, 16\penalty0 (8):\penalty0 T08001,
  August 2021{\natexlab{a}}.
\newblock \doi{10.1088/1748-0221/16/08/T08001}.

\bibitem[{Anfimov}(2017)]{2017JInst..12C6017A}
N.~{Anfimov}.
\newblock {Large photocathode 20-inch PMT testing methods for the JUNO
  experiment}.
\newblock \emph{Journal of Instrumentation}, 12\penalty0 (6):\penalty0 C06017,
  June 2017.
\newblock \doi{10.1088/1748-0221/12/06/C06017}.

\bibitem[{Petitjean} et~al.(2021){Petitjean}, {Clerbaux}, {Colomer Molla}, and
  {Yang}]{2021arXiv211012277P}
Pierre-Alexandre {Petitjean}, Barbara {Clerbaux}, Marta {Colomer Molla}, and
  Yifan {Yang}.
\newblock {The JUNO experiment and its electronics readout system}.
\newblock \emph{arXiv e-prints}, art. arXiv:2110.12277, October 2021.

\bibitem[{HVSys Company}({2013})]{HVSYS-LED}
{HVSys Company}.
\newblock {LIGHT EMITTING DIODE SOURCES OF CALIBRATED SHORT LIGHT FLASHES},
  {2013}.
\newblock http://hvsys.ru/images/data/news/10\_small\_1368803142.pdf.

\bibitem[{Anfimov} et~al.(2019){Anfimov}, {Rybnikov}, and
  {Sotnikov}]{2019NIMPA.939...61A}
N.~{Anfimov}, A.~{Rybnikov}, and A.~{Sotnikov}.
\newblock {Optimization of the light intensity for Photodetector calibration}.
\newblock \emph{Nuclear Instruments and Methods in Physics Research A},
  939:\penalty0 61--65, September 2019.
\newblock \doi{10.1016/j.nima.2019.05.070}.

\bibitem[{Bellato} et~al.(2021)]{2021NIMPA.98564600B}
M.~{Bellato} et~al.
\newblock {Embedded readout electronics R\&D for the large PMTs in the JUNO
  experiment}.
\newblock \emph{Nuclear Instruments and Methods in Physics Research A},
  985:\penalty0 164600, January 2021.
\newblock \doi{10.1016/j.nima.2020.164600}.

\bibitem[Zhou et~al.(2021)]{junoDAQ-prototype2021}
T.~Zhou et~al.
\newblock {DAQ} readout prototype for {JUNO}.
\newblock \emph{Radiat Detect Technol Methods}, 5:\penalty0 600–608, 2021.
\newblock ISSN 0168-9002.
\newblock \doi{https://doi.org/10.1007/s41605-021-00290-5}.
\newblock URL \url{https://doi.org/10.1007/s41605-021-00290-5}.
\newblock New Developments in Photodetection NDIP11.

\bibitem[{Wonsak} et~al.(2021{\natexlab{b}})]{JUNO-20inchPMT-testing}
B.~{Wonsak} et~al.
\newblock {A container-based facility for testing 20'000 20-inch PMTs for
  JUNO}.
\newblock \emph{Journal of Instrumentation}, 16\penalty0 (8):\penalty0 T08001,
  August 2021{\natexlab{b}}.
\newblock \doi{10.1088/1748-0221/16/08/T08001}.

\bibitem[Luo et~al.(2018)]{JUNOPMTsignalopt}
F.~Luo et~al.
\newblock {Signal Optimization with HV divider of MCP-PMT for JUNO}.
\newblock \emph{Springer Proc. Phys.}, 213:\penalty0 309--314, 2018.
\newblock \doi{10.1007/978-981-13-1316-5\_58}.

\bibitem[{Luo} et~al.(2016)]{2016ChPhC..40i6002L}
Feng-Jiao {Luo} et~al.
\newblock {PMT overshoot study for the JUNO prototype detector}.
\newblock \emph{Chinese Physics C}, 40\penalty0 (9):\penalty0 096002, September
  2016.
\newblock \doi{10.1088/1674-1137/40/9/096002}.

\bibitem[Zhang et~al.(2019{\natexlab{a}})]{waveAnalysisHaiqiong}
H.~Q. Zhang et~al.
\newblock {Comparison on PMT Waveform Reconstructions with JUNO Prototype}.
\newblock \emph{JINST}, 14\penalty0 (08):\penalty0 T08002, 2019{\natexlab{a}}.
\newblock \doi{10.1088/1748-0221/14/08/T08002}.

\bibitem[{CERN ROOT}(2020)]{cern-root-FFT}
{CERN ROOT}.
\newblock {Fast Fourier Transforms tutorials}, 2020.
\newblock URL \url{https://root.cern.ch/doc/master/group__tutorial__fft.html}.

\bibitem[Zhang et~al.(2022)]{zhangyu-large-pulse}
Yu~Zhang et~al.
\newblock {Study of 20-inch PMTs dark count generated large pulses}.
\newblock \emph{arXiv e-prints}, art. arXiv:2206.07456, July 2022.

\bibitem[Jetter et~al.(2012)Jetter, Dwyer, Jiang, Liu, Wang, Wang, and
  Wen]{dayabay-Jetter_2012}
Sören Jetter, Dan Dwyer, Wen-Qi Jiang, Da-Wei Liu, Yi-Fang Wang, Zhi-Min Wang,
  and Liang-Jian Wen.
\newblock {PMT} waveform modeling at the daya bay experiment.
\newblock \emph{Chinese Physics C}, 36\penalty0 (8):\penalty0 733--741, aug
  2012.
\newblock \doi{10.1088/1674-1137/36/8/009}.
\newblock URL \url{https://doi.org/10.1088/1674-1137/36/8/009}.

\bibitem[{Huang} et~al.(2018)]{Huang-waveform}
Yongbo {Huang} et~al.
\newblock {The Flash ADC system and PMT waveform reconstruction for the Daya
  Bay experiment}.
\newblock \emph{Nuclear Instruments and Methods in Physics Research A},
  895:\penalty0 48--55, July 2018.
\newblock \doi{10.1016/j.nima.2018.03.061}.

\bibitem[Zhang et~al.(2019{\natexlab{b}})]{Zhang_2019_waveform}
H.Q. Zhang et~al.
\newblock Comparison on {PMT} waveform reconstructions with {JUNO} prototype.
\newblock \emph{Journal of Instrumentation}, 14\penalty0 (08):\penalty0
  T08002--T08002, aug 2019{\natexlab{b}}.
\newblock \doi{10.1088/1748-0221/14/08/t08002}.
\newblock URL \url{https://doi.org/10.1088/1748-0221/14/08/t08002}.

\bibitem[Xu et~al.(2021)]{xu2021ultimate}
Dacheng Xu et~al.
\newblock Towards the ultimate pmt waveform analysis, 2021.
\newblock URL \url{https://arxiv.org/pdf/2112.06913.pdf}.

\end{thebibliography}

\end{document}